\documentclass[fleqn,usenatbib]{mnras}

\usepackage[T1]{fontenc}

\DeclareRobustCommand{\VAN}[3]{#2}
\let\VANthebibliography\thebibliography
\def\thebibliography{\DeclareRobustCommand{\VAN}[3]{##3}\VANthebibliography}

\usepackage{graphicx}	
\usepackage{amsmath}	
\usepackage{amssymb}	

\newcommand{\theth}{{\sc The Three Hundred}}

\usepackage{xcolor}

\title[Hydrostatic mass bias]{A study of the hydrostatic mass bias dependence and evolution within \theth\ clusters}

\author[G. Gianfagna et al.]{
Giulia Gianfagna,$^{1,2}$\thanks{E-mail: }
Elena Rasia,$^{3,4}$
Weiguang Cui,$^{5, 6}$
Marco De Petris,$^{2}$
Gustavo Yepes,$^{6,7}$
\newauthor Ana Contreras-Santos,$^{6}$ and
Alexander Knebe,$^{6,7,8}$ 
\\
$^{1}$INAF, Istituto di Astrofisica e Planetologia Spaziali, via Fosso del Cavaliere 100, 00133 Rome, Italy\\
$^{2}$Dipartimento di Fisica, Sapienza Universit\'a di Roma, Piazzale Aldo Moro, 5-00185 Roma, Italy\\
$^{3}$IFPU - Institute for Fundamental Physics of the Universe, Via Beirut 2, 34014 Trieste, Italy \\
$^{4}$INAF Osservatorio Astronomico di Trieste, via Tiepolo 11, I-34131, Trieste, Italy\\
$^{5}$Institute for Astronomy, University of Edinburgh, Royal Observatory, Edinburgh EH9 3HJ, UK\\
$^{6}$Departamento deF\'{\i}sica Te\'orica, M\'odulo 15, Facultad de Ciencias, Universidad Aut\'onoma de Madrid, E-28049 Madrid, Spain\\
$^{7}$Centro de Investigaci\'on Avanzada en F\'{\i}sica Fundamental (CIAFF), Facultad de Ciencias, Universidad Aut\'onoma de Madrid, 28049 Madrid, Spain\\
$^{8}$International Centre for Radio Astronomy Research, University of Western Australia, 35 Stirling Highway, Crawley, Western Australia 6009, Australia
}

\date{Accepted XXX. Received YYY; in original form ZZZ}

\pubyear{2021}
\usepackage{newtxtext,newtxmath}
\begin{document}
\label{firstpage}
\pagerange{\pageref{firstpage}--\pageref{lastpage}}
\maketitle

\begin{abstract}
We use a set of about 300 simulated clusters from \theth\ Project to calculate their hydrostatic masses and evaluate the associated bias by comparing them with the true cluster mass. Over a redshift range from 0.07 to 1.3, we study the dependence of the hydrostatic bias on redshift, concentration, mass growth, dynamical state, mass, and halo shapes. We find almost no correlation between the bias and any of these parameters. However, there is a clear evidence that the scatter of the mass-bias distribution is larger for low-concentrated objects, high mass growth, and more generically for disturbed systems. Moreover, we carefully study the evolution of the bias of twelve clusters throughout a major-merger event. We find that the hydrostatic-mass bias follows a particular evolution track along the merger process: to an initial significant increase of the bias recorded at the begin of merger, a constant plateaus follows until the end of merge, when there is a dramatic decrease in the bias before the cluster finally become relaxed again.
This large variation of the bias is in agreement with the large scatter of the hydrostatic bias for dynamical disturbed clusters. These objects should be avoided in cosmological studies because their exact relaxation phase is difficult to predict, hence their mass bias cannot be trivially accounted for.

\end{abstract}




\section{Introduction}

Galaxy clusters are the most massive gravitationally-bound structures in the Universe and, as such, their mass plays a key role in the estimation of the cosmological parameters. The majority of their mass is composed by Dark Matter (DM, almost 80\%), the remaining part is composed by stars (galaxies) and hot gas which is diffused between the galaxies and it is called Intra-Cluster Medium, ICM \citep[see][for a review]{bib:kravtsov_rev}. 

The mass of these objects in observations can be estimated in several ways: from galaxy kinematic, applying the virial theorem to the distribution of cluster galaxies \citep[for example][]{Li2021}, or directly from the velocity distribution of cluster galaxies \citep{Zwicky1937, Pratt2019, Tian2021, hernandezlang2021}; from scaling laws, linking their total mass to several observable quantities, like X-rays luminosity or SZ (Sunyaev-Zeldovich) brightness \citep{Nagarajan2018, Bulbul2019}; from weak lensing \citep{vonderLinden2014, Hoekstra2015, Okabe2016, Artis2021} and, recently, also combining the mentioned methods and using machine learning \citep{Ntampaka2015, deandres2021}. Another frequently used approach is to derive the cluster mass under the assumption of the hydrostatic equilibrium (HE). This method uses temperature, density and pressure profiles of hot gas extracted from X-rays alone or combined with SZ effect observations (see \citealt{Pratt2019} for a recent review).
Three assumptions are the basis of the HE: the ICM must trace the cluster potential well, it must be spherically symmetric, and its pressure must be purely thermal. 
As shown in a recent review by \citet{gianfagna2021}, the HE mass evaluated in simulated samples is on average underestimating the true mass. In this paper we investigate more this subject and in particular the dependence of the bias by several other parameters that characterise the cluster dynamical state.

Numerical simulation studies find that the bias has a negligible dependence with the redshift \citep{piffaretti, lebrun, Henson2017, gianfagna2021}. On the contrary, observational data are consistent with a decreasing mass bias along the redshift, see for instance \cite{Salvati2019} and \cite{wicker2021} on \textit{Planck} clusters. This dependence could be linked to a mass dependence of the bias, as clusters observed at higher redshifts tend to have a higher mass. However, the mass dependence is not detected in simulations, either using the spectroscopic-like weighted temperature\footnote{Note that spectroscopic-like weighted temperature is more similar to the X-ray temperature obtained from X-ray spectroscopic analysis of observed data.} \citep{mazzotta2004} to estimate the HE mass \citep{piffaretti, lebrun, Henson2017, pearce2019, ansarifard, barnes2020} or the mass-weighted temperature. Several authors also study the dependence of the bias on the dynamical state, the general findings also pointed towards no correlation, even if there is a hint that disturbed objects have a large scatter with respect to the relaxed ones \citep{piffaretti, rasia, Nelson2014, Henson2017, ansarifard}. 
By comparing results in literature it appears that the hydrostatic mass bias does not depend on various cosmological-simulation ingredients, such as the baryon physics included in the simulations, the mass particles resolution or the number of objects in the samples \citep{gianfagna2021}. Due to this consistency for simulated results, the discrepancy on the bias dependence on redshift between simulation predictions and observation results is still puzzling. This could be due to less numerous observational samples, due to a different mass sampling at high redshift and due to the uncertainty in the total physical mass estimation in observation.

The selection may also play a key role, for example the inclusion of disturbed objects. Indeed, all of the HE assumptions are violated during major mergers events which
can induce strong dynamical perturbations, adding non-thermal support to the equilibrium budget \citep{Nelson2012, rasia, Rasia2013, Angelinelli2020, Sereno2021}. 
Therefore, the hydrostatic bias can significantly deviate from the expectation. 
In post-merger situations, then the merging object might have already been destroyed and thus it cannot be distinguished as a separate structure. 
Applying the hydrostatic-equilibrium technique to these cases (assuming they are in a relaxed state) will lead to a misinterpretation. 
Studying the extreme cases in simulations, where the dynamical state is clearly identified, will allow us to better understand the situations when the HE assumption is strongly violated and how the HE mass bias is impacted by mergers.

The organisation of this paper is as follow: in Section \ref{sec:300} we introduce \theth\ set of galaxy clusters, in Section \ref{sec:HE} we give an overview on the hydrostatic equilibrium method. In Section \ref{sec:methods} the ICM radial profiles are introduced, along with their analysis, and we introduce the parameters which will be used to characterise the dynamical state, the accretion and the geometry of clusters. In Section \ref{sec:results} we provide our the results, and we conclude in Section \ref{sec:conclusions}.


\section{\theth\ Project}
\label{sec:300}

This work is based on clusters simulated for \theth\ Project, introduced in \citet{Cui2018}.
This project is based on the resimulations of a set of 324 Lagrangian regions extracted from the MultiDark Planck (MDPL2) dark-matter-only (DM) cosmological simulation  \citep{Klypin2016} with different simulation codes. MDPL2 consists of a periodic cube of comoving length $1h^{-1}$ Gpc and $3840^3$ N-body particles.
The Lagrangian regions are identified at $z = 0$ around the most massive objects found in the parent simulation with a comoving radius of $15h^{-1}$ Mpc. The particles in those massive clusters and in their surrounding were traced back to $z=120$ to generate the initial conditions for the zoomed-in re-simulations.
The original mass resolution was kept in the central region, and those particles were split into dark matter and gas particles based on the cosmological baryon fraction $\Omega_b$. The high-resolution dark-matter-particle mass is $1.87\times10^{9} \rm M_{\odot}$, while the initial gas mass is $3.48\times10^8 \rm M_{\odot}$.  
In the outer region, the dark matter particles were degraded with multiple levels of mass refinements in several concentric shells.
The 324 zoomed-in regions have been hydro-dynamically re-simulated with three different baryon models: GADGET-X \citep{Rasia2015}, GADGET-MUSIC \citep{Sembolini2013} and GIZMO-Simba \citep{Dave2019,Cui2022}. In this work we use the GADGET-X simulated clusters. This code includes several radiative processes, like gas cooling, star formation, thermal feedback  from  supernovae, chemical evolution and enrichment, super massive black holes with AGN feedback \citep[see][for details]{Cui2018}.

The simulations assume a standard cosmological model according to the \citet{Planck2016} results: $h = 0.6777$ for the reduced Hubble parameter, $n = 0.96$ for the primordial spectral index, $\sigma_8 = 0.8228$ for the amplitude of the mass density fluctuations in a sphere of $8 h^{-1}$ Mpc comoving radius, $\Omega_{\Lambda} = 0.692885$, $\Omega_m = 0.307115$, and $\Omega_b = 0.048206$ respectively for the density parameters of dark energy, dark matter, and baryonic matter.

\subsection{Sample}

For the analysis presented here, we select the most massive cluster for each region that does not include any low-resolution particles within its virial radius. 
The evolution of the bias is tested at 9 redshifts, covering the redshift range between $z=0.07$ and $z=1.32$.  
The mean mass and the number of objects considered at each redshift are reported in Table \ref{Tab:z_N_mass} for the three studied overdensities: $\Delta=$ 2500, 500 and 200.  
These indicate the radius $R_{\rm \Delta}$ of a sphere whose density is either 2500, 500 or 200 times the critical density of the Universe at the consider cosmic time:
\begin{equation}
\rho_{\rm crit} = (3/8\pi G) H_0^2 [\Omega_M (1+z)^3 + \Omega_{\Lambda}],
\label{eq:rho_crit}
\end{equation}
where $H_0$ is the Hubble constant and G the gravitational constant. 

\begin{table}
\centering
\caption{The redshifts analysed in this work are written in the first column. 
The number of objects 
are in the second column. In the third, fourth and fifth columns are the mean masses at the three nominal overdensities 2500, 500, 200 with units of $10^{14} M_{\odot}$.}
\begin{tabular}[t]{ccccc}
\hline
 $z$  &  $N$ & $<M_{2500}>$ & $<M_{500}>$ & $<M_{200}>$ \\
  & & $\rm [10^{14} M_{\odot}]$& $\rm [10^{14} M_{\odot}]$& $\rm [10^{14} M_{\odot}]$\\
\hline
 1.32 & 277 & 0.47 & 1.36 & 1.99 \\ 
 1.22 & 281 & 0.55 & 1.58 & 2.31 \\
 0.99 & 304 & 0.78 & 2.26 & 3.29 \\
 0.78 & 305 & 1.07 & 3.04 & 4.45 \\
 0.59 & 305 & 1.48 & 4.05 & 5.91 \\
 0.46 & 300 & 1.73 & 4.89 & 7.22 \\
 0.33 & 297 & 2.09 & 5.73 & 8.46 \\
 0.22 & 298 & 2.43 & 6.73 & 9.97 \\
 0.07 & 290 & 2.98 & 8.19 & 12.20 \\
\hline
\end{tabular}
\label{Tab:z_N_mass}
\end{table}

\section{The hydrostatic equilibrium mass}
\label{sec:HE}

As said in the introduction, the hydrostatic equilibrium hypothesis is often at the basis of the procedure that lead to estimate the mass of galaxy clusters from X-ray observations \citep{bib:kravtsov_rev, bib:ettori_rev}.
This assumption foresees that the gas thermal pressure, which naturally leads to expansion of the gas, is balanced by the gravitational forces, which, on the other side, causes the systems to collapse. The equilibrium is expressed as the equality between the gradient of the thermal pressure and that of the gravitational potential and it is supposed to hold at every concentric distance from the cluster centre. 
Connecting the gravitational potential to the mass, under the spherical assumption, leads to the following formula for the total mass inside a sphere of radius $r$:
\begin{equation}
M_{\rm HE, SZ}(<r) = -\frac{r^2}{G \rho_{\rm g}(r)} \frac{{\rm d} P_{\rm th}(r)}{{\rm d}\ r}
\label{eq:Mhe_P}
\end{equation}
where 
$\rho_{\rm g}$ and $P_{\rm th}$ are the density and the thermal pressure of the gas, respectively. We will refer to this mass as $M_{\rm HE, SZ}$, since the radial pressure profile can be derived from Compton-$y$ parameter maps provided by clusters SZ observations at millimetre wavelengths.

Assuming the equation of state of an ideal gas, the cluster mass can also be estimated with gas density and temperature separately: 
\begin{equation}
M_{ \rm HE, X}(<r) = - \frac{rk_{\rm B} T(r)}{G\mu m_{\rm p}}\left[ \frac{{\rm d}\ln \rho_{\rm g} (r)}{{\rm d}\ln r} + \frac{{\rm d}\ln T(r)}{{\rm d}\ln r}\right].
\label{eq:Mhe_T}
\end{equation}
We refer to this HE mass formulation as $M_{\rm HE, X}$ because historically this expression was used in the analysis of X-rays observations \citep{ansarifard}.
The detailed computation of the HE mass in our simulated sample will be introduced in next section.

This mass is compared with the true total mass of the systems, obtained by summing over all dark matter, stars and gas particle masses inside a fixed aperture radius. The mass bias $b_{\rm SZ}$ or $b_{\rm X}$
is defined as
\begin{equation}
\label{eq:bias}
b = \frac{M_{\rm true} - M_{\rm HE}}{M_{\rm true}}.
\end{equation}
This bias can be either positive, when the HE mass is underestimating the true cluster mass, or negative, when there is an overestimation of the true mass. A null bias results in a perfect estimate of the true mass through HE.

\section{Methods}
\label{sec:methods}

In this work, we aim to estimate the two hydrostatic masses, as presented in Eqs.~(\ref{eq:Mhe_P}) and (\ref{eq:Mhe_T}).
The ICM temperature, gas density and pressure radial profiles play a fundamental role in the calculation of these masses. In this section, we describe how we compute them in \theth\ simulations together with other relevant quantities which we use to study the correlations with the bias. The quantities linked to the study of the mass bias during a major merger event will be described in Section~\ref{sec:merger}.

\subsection{The ICM profiles}
\label{subsection:radial_profiles}

The cluster profiles are extracted in logarithmically equal-spaced radial bins, centred at the maximum density (mostly coinciding with the minimum of the potential well, see also \citealt{Cui2016,ansarifard}). Only gas particles with density below the star-formation density threshold and with temperature above 0.3 keV are considered as ICM for calculation, this is to insure that we are selecting the hot particles which can generate X-ray or SZ signals, in order to match the observations. 

Once the profiles are extracted, we fit them with analytic models which will be described below.
We then use the analytical best-fitting curves to estimate the hydrostatic masses. 
Following this strategy will allow to avoid the small fluctuations that are presented in the 3D numerical profiles which strongly impact the calculation of the derivatives in the hydrostatic-equilibrium mass equations \citep[see][]{gianfagna2021}. On average, this still allow to well fit the large fluctuations (bumps or deeps in the profiles), caused by substructures, which should instead taken into account. This is valid especially for the temperature and density models, that have the largest number of parameters. The pressure profile model could instead cancel some large fluctuation, due to the smaller number of parameters, leading to a non-reliable SZ bias, but the two biases are compatible between each other (see Section \ref{sec:results}) and also the scatter is not significantly different.
The best-fit procedure of each profile is performed in the radial range [0.2-3]$\rm R_{500}$. This range was chosen to include the three studied overdensities: $\Delta=$ 2500, 500 and 200. We study the goodness of the fits using the $\chi^2$ test and found that all the fits can be classified as good fits.

\subsubsection{Gas Density}
The 3D gas density profiles are estimated as the total gas mass in a spherical shell, divided by the shell volume: $\rho = \sum m_{\rm i}/V_{\rm shell}$. The model chosen to fit this profile is the one proposed by \citet{vikh}:
\begin{equation}
\rho_g(r) = \rho_0^2 \frac{(r/r_{\rm d})^{-a}}{ (1 + (r/r_{\rm d})^2)^{3b-a/2}} \frac{1}{ (1 + (r/r_{\rm s})^c)^{e/c} } + 
\label{eq:rho_vikh}
\end{equation}
\begin{equation*}
\hspace{1cm} + \frac{\rho_{02}^2}{ (1 + (r/r_{\rm d2})^{2})^{3b_2} },
\end{equation*}
where we fix the parameter $c$ equal to 3 and we impose the following limitation: $e < 5$. The other 8 parameters are left free. Often in literature this model is simplified by discarding the second beta model, here we keep it to have a reliable fit also in the cluster region near $\rm R_{2500}$. For reference, we report in Table~\ref{Tab:vikh_rho} the medians of all best-fit parameters computed from the sample at $z=0.07$.

\begin{table}
\centering
\caption{
The medians and $16^{\rm th} - 84^{\rm th}$ percentiles of best-fit free parameters in Eq.\ref{eq:rho_vikh} for gas density profiles and the reduced $\chi^2$. The results only for all $z=0.07$ clusters are exhibited here. The $c$ parameter is fixed to 3.}
\begin{tabular}[t]{ccccc}
\hline
   $\rho_0$[$10^{13} \rm M_{\odot}/Mpc^3$] & $r_d$[Mpc] & $r_s$[Mpc] & $a$\\
 $3.5^{+1.5}_{-1.3}$ & $0.71^{+0.20}_{-0.17}$ & $0.73^{+0.65}_{-0.29}$ & $1.2^{+1.2}_{-0.9}$ \\
\hline
$b$ & $e$ & $\rho_{0,2}$[$10^{13} \rm M_{\odot}/Mpc^3$] & $r_{d,2}$[Mpc]\\
$2.6^{+0.4}_{-0.7}$ & $2.5^{+0.4}_{-0.6}$ & $2.4^{+0.6}_{-0.8}$ & $0.95^{+0.36}_{-0.28}$ \\
\hline
$b_2$ & $\chi^2$ \\
$1.2^{+0.3}_{-0.2}$ & $1.1^{+0.8}_{-0.4}$\\
\hline
\end{tabular}
\label{Tab:vikh_rho}
\end{table}

\subsubsection{Gas Temperature}
The mass-weighted temperature profile is estimated as a weighted average over the gas particles in the same spherical shells used for the gas density:
\begin{equation}
T = \frac{\sum_i (m_i T_i)}{\sum_i m_i}.
\end{equation}
where $m_i$ and $T_i$ are the mass and temperature of $i^{th}$ gas particle. Using the mass-weighted temperature is the optimal choice for hydrostatic equilibrium studies which theoretically connect the temperature with the gravitational mass \citep{Biffi2014} and it is favorable with respect to the spectroscopic-like temperature \citep{mazzotta2004}. 
The analytical model for the mass-weighted temperature used in this work was introduced again by \citet{vikh}:
\begin{equation}
T(r) = T_0 \ \frac{x + \tau}{x+1} \ \frac{(r/r_{\rm t})^{-\alpha}}{ (1 + (r/r_{\rm t})^\beta)^{\gamma/\beta} },
\label{eq:T_vikh}
\end{equation}
\begin{equation*}
x = (r/r_{\rm cool})^{\alpha_{\rm cool}}.
\end{equation*}
All the 8 parameters are free to vary. 
Their medians for the $z=0.07$ clusters are reported in Table~\ref{Tab:vikh_t}.

\begin{table}
\centering
\caption{The medians and $16^{\rm th}$ and $84^{\rm th}$ percentiles of best-fit free parameters of Eq.\ref{eq:T_vikh} and the reduced $\chi^2$ computed for all $z=0.07$ clusters.}
\begin{tabular}[t]{cccccc}
\hline
   $T_0$[keV] & $\tau$ & $r_t$[Mpc] & $\alpha$ & $\beta$ & $\gamma$ \\
 $15.3^{+56.4}_{-10.3}$ & $0.37^{+0.32}_{-0.33}$ & $2.3^{+22.6}_{-1.29}$ & $0.29^{+0.71}_{-0.29}$ & $4.6^{+7.4}_{-4.6}$ & $1.3^{+2.2}_{-1.3}$ \\
\hline 
 $r_{cool}$[Mpc] & $\alpha_{cool}$ & $\chi^2$ & & & \\
 $1.4^{+1.5}_{-0.7}$ & $5.4^{+4.6}_{-2.8}$ & $1.4^{+0.4}_{-0.1}$ \\
\hline

\end{tabular}
\label{Tab:vikh_t}
\end{table}

\subsubsection{Gas Pressure}
\label{subsubsec:P}

The radial thermal pressure profile is measured from the gas density and temperature of each gas particle. It is 
modelled by the generalised Navarro-Frenk-White (gNFW) model \citep{nagai}:
\begin{equation}
P(r) = \frac{P_0}{x^{i} (1 + x^{g})^{\frac{h - i}{g}}},
\label{eq:gNFW}
\end{equation}
where $x = r/r_s$ is a dimensionless radial distance normalised to the scale radius, $r_s$.
The parameters $h$ and $i$ are the slopes for outer and inner region, respectively, and $g$ is the steepness of the transition between the two regimes. This model has 4 free parameters, since $i = 0.31$ is fixed \citep{arnaud10}. The medians of the best-fit parameters for the $z=0.07$ sample are listed in Table~\ref{Tab:gnfw}. 

\begin{table}
\centering
\caption{The medians and $16^{\rm th}$ and $84^{\rm th}$ percentiles of best-fit free parameters of Eq.\ref{eq:gNFW} and the reduced $\chi^2$ computed for all $z=0.07$ clusters. The $i$ parameter is fixed equal to $0.31$.}
\begin{tabular}[t]{ccccc}
\hline
   $P_0 [10^{-2} \rm keV/cm^3]$ & $r_s$ [Mpc] & $g$ & $h$ & $\chi^2$ \\
$2.6^{+5.5}_{-1.5}$ & $3.1^{+6.9}_{-2.5}$ & $1.1^{+1.3}_{-0.4}$ & $8.2^{+6.8}_{-4.0}$ & $1.1^{+0.8}_{-0.3}$\\
\hline

\end{tabular}
\label{Tab:gnfw}
\end{table}

\subsection{The NFW concentration}
Similarly to the thermodynamical profiles, we compute the total density profiles which include the contribution from all particles in the simulations (dark matter, gas, stars). The total mass profile is fitted with a NFW model \citep{nfw}:
\begin{equation}
M(<r)= M_0 \left[ \log(1+x) -\frac{x}{1+x} \right]. 
\end{equation}
As before the radial coordinate is related to the scale radius, $x=r/r_s$. This parameter, together with the normalisation, is derived from a best-fitting procedure applied in the radial range between 100 kpc and $R_{200}$, which roughly reproduce a typical radial range used in weak-lensing analysis. The concentration within R$_{500}$ is defined from the scale radius: $c_{500}=R_{500}/r_s$. 

\subsection{Relative mass growth}
We straightforwardly define the cluster mass growth as the relative mass difference between an initial, $t_{1}$, and a final time, $t_{2}$:
\begin{equation}
\frac{\Delta M/M} {{\rm d} t} = \frac{(M(t_2) -M(t_1))/M(t_1)}{[t_2-t_1]}.   
\end{equation}
Specifically, we estimated the mass growth from the merger tree between $z_2=0.33$ and $z_1=0.46$, so that $t_2-t_1$ corresponds to about 1 Gyr (precisely it is equal to $1.046$ Gyr). This interval corresponds to 5 simulation snapshots.

\subsection{Dynamical state parameter}
\label{subsec:dynamical_state}

The dynamical state of the clusters has been inferred by the combination of two indicators computed in 3D \citep{neto, cui2017, Cialone, DeLuca}:
\begin{itemize}
    \item $f_s=M_{\rm sub}/M_{\Delta}$, which is the fraction of cluster mass included in substructures inside a fixed overdensity;
    \item $\Delta_{\rm r} = |\textbf{r}_{\delta} - \textbf{r}_{\rm cm}|/R_{\Delta}$, which is the offset between the positions of the maximum density peak and the centre of mass of the cluster within $R_{\Delta}$ and normalised to that aperture radius. 
\end{itemize}

When correlating the dynamical-state and the HE mass bias, we evaluate all quantities of interest at the same overdensity value. 
In order to have a relaxed cluster, both indicators should be smaller than 0.1 \citep{Cialone} while they should be both greater than 0.1 to have a disturbed one. The other cases are classified as intermediate or hybrid. The percentage of relaxed clusters is almost 50\% at low redshifts, and decreases to 30\% at high redshifts, while the percentage of disturbed clusters is 20\% at low redshifts and increases to 30\% at high redshifts \citep{DeLuca}.
As previously said, in our study we consider one combined parameter as in \cite{Haggar2020} but derived exclusively from the two mentioned indicators as in \cite{DeLuca}:

\begin{equation}
    \chi_{\rm DS} = \sqrt{\frac{2}{\left(\frac{\Delta_{\rm r}}{0.1}\right)^2 + \left( \frac{f_s}{0.1}\right)^2 
    }} .
\end{equation}
Using the $\chi_{\rm DS}$ parameter is a continuous way to classify the state of the cluster where the relaxed ones satisfy $\chi_{\rm DS}>1$. We use this parameter to study the dependence of the dynamical state on the HE bias (see Section \ref{subsec:relaxation_p}). An extensive study of the relaxation state of \theth\ clusters including its dependence from redshift is presented in \citet{DeLuca}.

\subsection{Triaxiality}
In order to test how much the lack of spherical symmetry is affecting the HE mass estimate, we calculate, using all particles within $R_{500}$, the three axes of the inertia tensor as in \citep{vega}. We then use the ratio $c/a$ and the triaxiality of the halos:
\begin{equation}
    t = \frac{1 - (b/a)^2}{1 - (c/a)^2},
\label{eq:triaxiality}
\end{equation}
where $a, b$ and $c$ indicate the major, intermediate, and minor axes, respectively. Depending on the value of the parameter $t$, the halo is considered as oblate if $0 < t < 1/3$; triaxial if $1/3 < t < 2/3$ and prolate if $2/3 < t < 1$. More simply, when $c/a$ is equal to $1$ the cluster is spherical.

\section{Results}
\label{sec:results}
In this section we present the results of our work. First, in Section~\ref{sec:bias_corr}, we study the HE mass bias dependencies on the redshift, mass, and on all other parameters: the concentration, the mass growth, the relaxation parameter and the triaxiality. 
We performed all these analyses at the three overdensities  ($\Delta=2500,500,200$) and at all considered redshifts. However, most of the figures related to this part of the work (with the exception of the first) are only presented with $\Delta=500$ and at one specific redshift because we do not find a strong dependence of the results on either of these two quantities.

Secondly, in Section~\ref{sec:merger}, we analyse how the HE mass bias varies throughout strong merger events. In this case, we only analyse the behaviour of the bias at $\rm R_{200}$, since we take as reference the mergers analysis in \cite{Contreras2022}, which is done considering the entire cluster volume and thus at overdensity 200.

\subsection{Bias correlations}
\label{sec:bias_corr}

\subsubsection{Redshift dependence}

\begin{figure}
    \includegraphics[width=.5\textwidth]{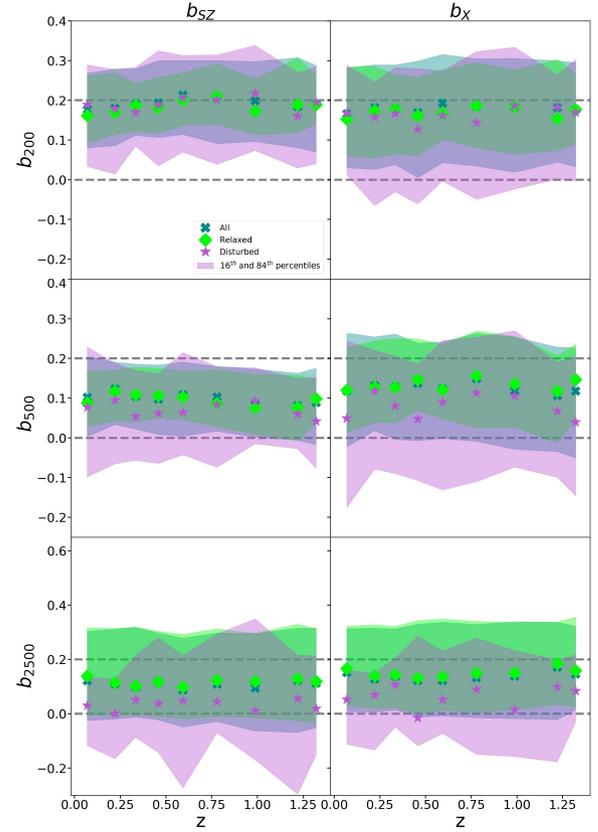} 
    \caption{The redshift evolution of the biases, $b_ {\rm SZ}$ (left panels) and $b_{\rm X}$ (right panels). The median values of the bias for all clusters, relaxed and un-relaxed clusters are represented with dark cyan crosses, green diamonds and purple stars respectively. The shaded regions represent the $16^{th}$ and $84^{th}$ percentiles. The bias estimated at $\rm R_{200}$, $\rm R_{500}$ and $\rm R_{2500}$ are represented in the top, middle and bottom panels respectively. The dashed lines show the 0 and 0.2 bias for reference.}
 \label{fig:redshift}
\end{figure}

The variation of $b_{SZ}$ and $b_X$ along the redshift range is presented in Fig. \ref{fig:redshift} at the 3 overdensities and for the different cluster dynamical states defined with the dynamical-state parameter, $\chi_{\rm DS}$.  For all overdensities, the un-relaxed clusters (purple shaded region) have the largest scatter. Interestingly, this large scatter is mostly caused by the presence of low-value bias. We will present the reason in Section~\ref{sec:merger}. 
No dependence of the bias on the redshift up to $z = 1.25$ is detected in agreement with  \citealt{lebrun, Henson2017, salvati, Koukoufilippas2020}. 
Furthermore, we notice that the median $b_{2500}$ is very similar to the median $b_{500}$ and that both are close to 0.1 (10\% of bias) and as such systematically lower than $b_{200}$ (close to $\sim 0.2$). This is in agreement with \cite{gianfagna2021}, which showed a declining $b$ from outer to inner radius. This is occurring despite the differences both in the code (GADGET2 versus GADGET-X) and in the baryon models. Note, however, that the SZ and X median $b_{500}$ in \cite{gianfagna2021} is slightly larger than this work. The biases dispersion at $R_{2500}$ seems to be slightly larger with respect to the other overdensities, this is probably due to the larger deviations in the cluster core properties, which could marginally affect the profile at $R_{2500}$. In \theth, the simulation produce a more diverse variety of simulated cores with respect to the GADGET2 code, and in some situations, the simulated clusters are extremely peaked at the centre \citep[see][]{campitiello2022}.


\subsubsection{Concentration}
The concentration parameter $c_{500}$ of a halo is representative of the halo's central density. In presence of a disturbed system the concentration is typically lower since the X-ray peak might have been destroyed by a merger event which also could have brought more mass in the external regions. For this it might be interesting to see if there is any correlation between the HE mass bias and the NFW concentration parameter.

\begin{figure}
    \includegraphics[width=.5\textwidth]{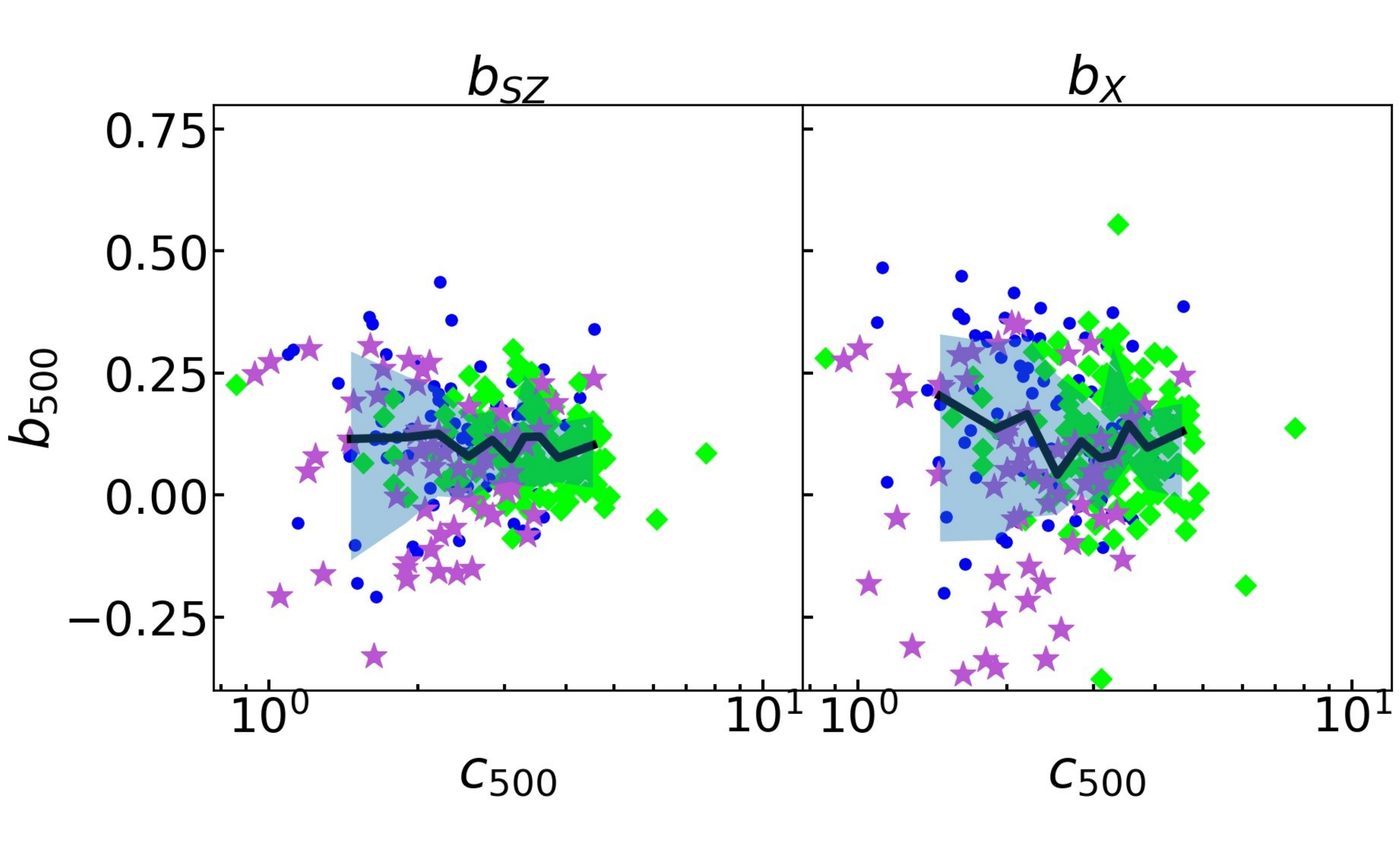}
    \caption{The bias dependence on the concentration at $\rm R_{500}$ and at $z = 0.07$, $b_ {\rm SZ}$ is represented in the left panel, $b_{\rm X}$ in the right one. The black line represents the median bias value, with the shaded regions as 16th and 84th percentiles. The relaxed clusters are shown in green diamonds, the disturbed in purple stars and the hybrids in blue dots.}
 \label{fig:concentration}
\end{figure}

The bias as a function of the concentration is represented in Fig. \ref{fig:concentration} where the relaxed, disturbed and hybrid clusters are introduced with different symbols and colours. 
Hybrid clusters are these with either $f_s <0.1$ or $\Delta_r<0.1$. The two quantities do not show any dependence, as clear from the trend of the median value shown with a black line. We notice however that the scatter in the bias substantially decreases going towards larger concentration values. This is quantified by the 
half difference between the $16^{\rm th}$ and the $84^{\rm th}$ percentiles, $d2=0.5 \times (p_{84} - p_{16})$ reported in Table~\ref{Tab:concentration} together with the bias percentiles. The value of $d2$ decreases from low (third row) to high concentrated clusters (first row) by 70-80\%. 
This trend -- reduced scatter with increasing concentration -- is expected because most of the clusters with higher concentration are relaxed.

\begin{table}
\centering
\caption{Table of the biases for the 50 clusters with the highest (first row) and lowest (third row) NFW concentration and the remaining
clusters (second row) for the $z=0.59$ 0 sample. We report the median value ($m$), the $16^{\rm th}$ and $84^{\rm th}$ percentiles ($p_{16}$ and $p_{84}$), and their half difference ($d_2$) at $R_{500}$.
}
\begin{tabular}[t]{ccccccccc}
\hline
$c_{500}$   & $b_{SZ}$ & & & & $b_X$ & \\
  & $m$ & $p_{16}$ & $p_{84}$ & $d_2$ & m & $p_{16}$ & $p_{84}$ & $d_2$ \\
\hline
high & 0.11 & 0.07 & 0.19 & 0.06 & 0.12 & 0.05 & 0.24 & 0.10 \\
\hline
med & 0.11 & 0.00 & 0.18 & 0.09 & 0.13 & 0.00 & 0.24 & 0.12 \\
\hline 
low & 0.06 & -0.03 & 0.22 & 0.12 & 0.11 & -0.07 & 0.24 & 0.15 \\

\hline

\end{tabular}
\label{Tab:concentration}
\end{table}

%

\subsubsection{Relative Mass growth}
\begin{figure}
    \includegraphics[width=.5\textwidth]{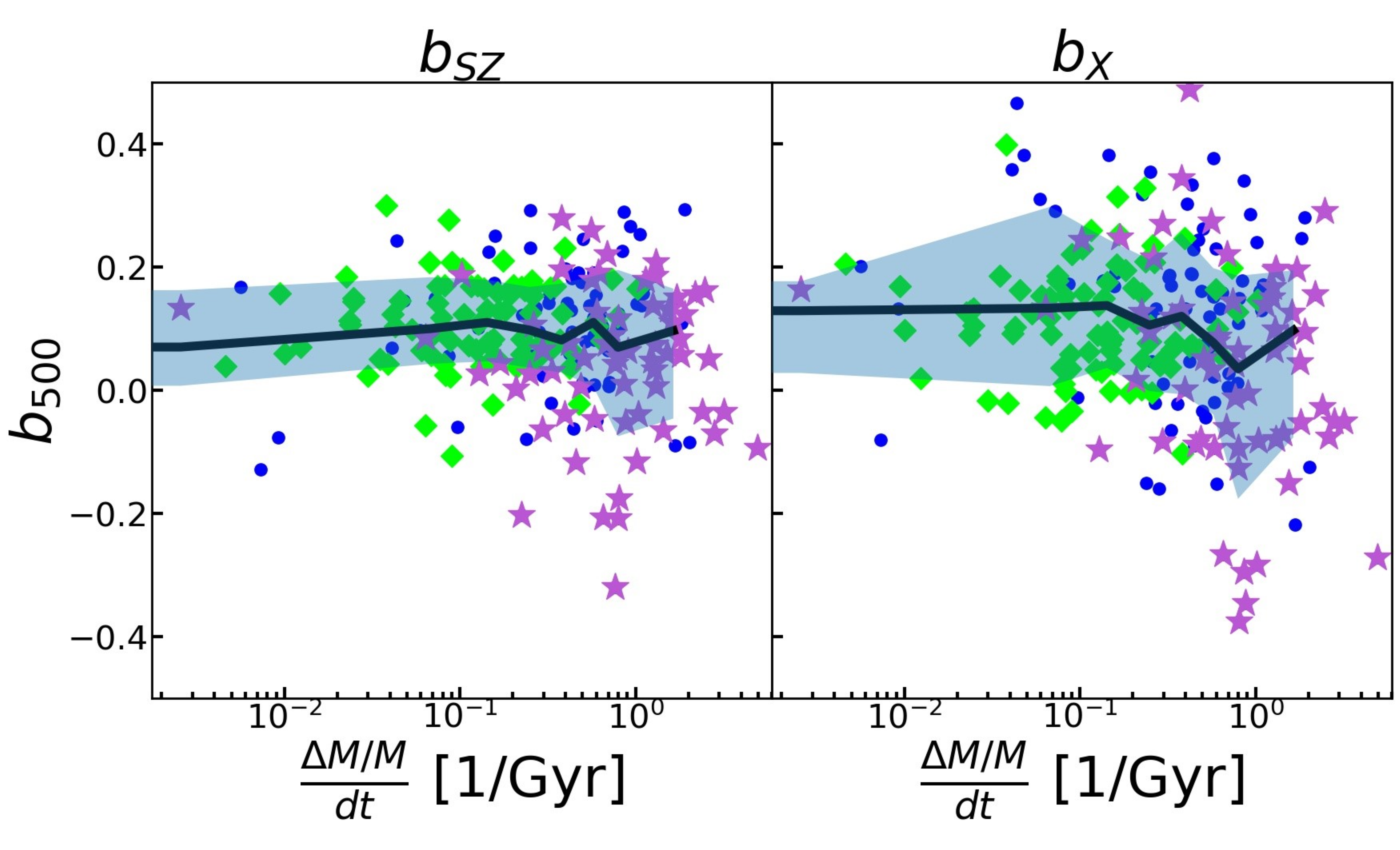}
    \includegraphics[width=.5\textwidth]{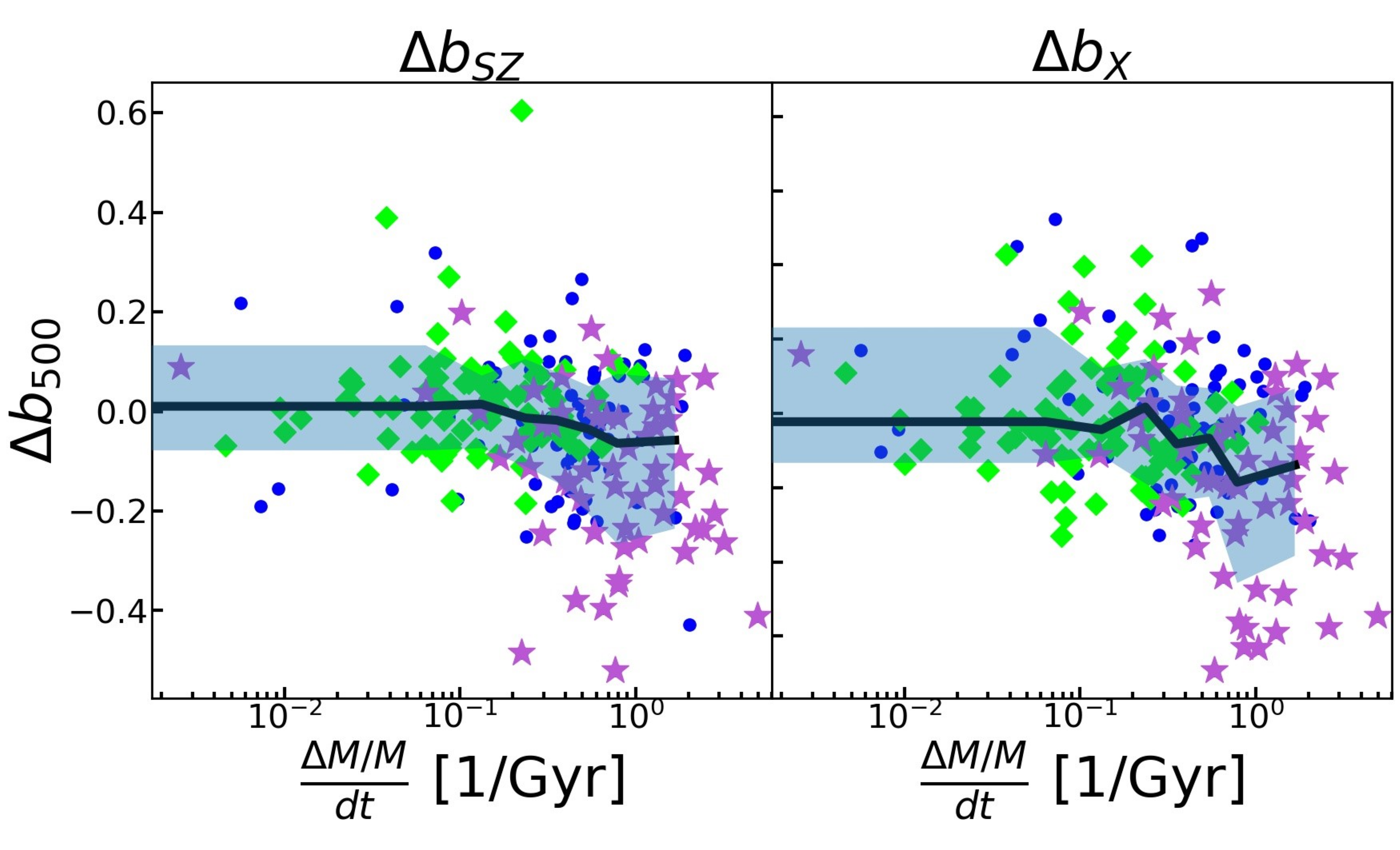}
    \caption{Top panel: the bias dependence on the mass growth $\frac{\Delta M/M}{dt}$ is represented. $\frac{\Delta M/M}{dt}$ is estimated from $z=0.46$ to $z=0.33$, which corresponds to 1 Gyr, while the biases are estimated at $z=0.33$. Bottom panel: the bias variation from $z=0.46$ to $z=0.33$ as a function of $\frac{\Delta M/M}{dt}$. The relaxed and disturbed clusters are represented with green diamonds and purple stars respectively, while the hybrids with blue dots. The cluster dynamical states shown in this plot are estimated at $z=0.33$. }
 \label{fig:accretion_rate}
\end{figure}

The dependence of the bias at $z=0.333$ on the mass growth recorded in the last Gyr (and specifically since $z=0.46$) 
is shown in the top panel of Fig. \ref{fig:accretion_rate}, where both quantities are estimated at $\rm R_{500}$. The relative median values for high, low, and intermediate accreting clusters are listed in Table~\ref{Tab:mass_growth}.
The correlation is again very weak, the Pearson correlation coefficient is -0.20 for $b_{\rm SZ}$ and -0.36 for $b_{\rm X}$. However, the systems with the largest mass growth have also the largest dispersion. This is expected because the systems with the largest mass growth are also the most disturbed (purple stars in Fig.~\ref{fig:accretion_rate}). This relation will be discussed more afterwards, however, we point out that clusters with large mass accretion will also have abundant non-thermal gas motions and, thus, a large non-thermal pressure component. 
The same trend is seen for the quantities estimated at $\rm R_{200}$, with a Pearson correlation coefficient of 0.16 for $b_{\rm SZ}$ and 0.21 for $b_{\rm X}$. 

In the bottom panel of Fig.  \ref{fig:accretion_rate} the difference between the bias at $z_2=0.33$ and at $z_1=0.46$ is shown ($\Delta b=b(z_2)-b(z_1)$). The variation in the bias for disturbed clusters (at large accretion rates) is negative, implying that the HE mass bias overall decreases. This phenomenon which might seem surprising will be better studied in Section~\ref{sec:merger}, but we can already see from the top panel the origin of this trend: after a major merger (with a mass variation of at least 50\% $\Delta M/M> 0.5$) the mass computed by Eqs.~\ref{eq:Mhe_P} or \ref{eq:Mhe_T} can be significantly larger than the true mass with a large scatter as well \citep[see also][]{Nelson2012}.


\begin{table}
\centering
\caption{Table of the biases values for the 50 clusters with the largest mass growth (first row), the 50 with the smallest (last row) and the remaining clusters (second row) for the $z=0.33$ sample. We report the median value ($m$), the $16^{\rm th}$ and $84^{\rm th}$ percentiles ($p_{16}$ and $p_{84}$), and their half difference ($d_2$) at R$_{500}$.
}
\begin{tabular}[t]{ccccccccc}
\hline
Mass    &    $b_{SZ}$ & & & & $b_X$ & \\
growth   &m & $p_{16}$  & $p_{84}$ & d2 & m & $p_{16}$ & $p_{84}$ & d2 \\
\hline
high & 0.09 & -0.05 & 0.17 & 0.11 & 0.09 & -0.12 & 0.20 & 0.16 \\
\hline
med & 0.10 & -0.02 & 0.17 & 0.08 & 0.12 & 0.00 & 0.22 & 0.11 \\
\hline 
low & 0.14 & 0.06 & 0.22 & 0.08 & 0.24 & 0.10 & 0.35 & 0.13 \\
\hline
\end{tabular}
\label{Tab:mass_growth}
\end{table}

\begin{figure}
    \includegraphics[width=.45\textwidth]{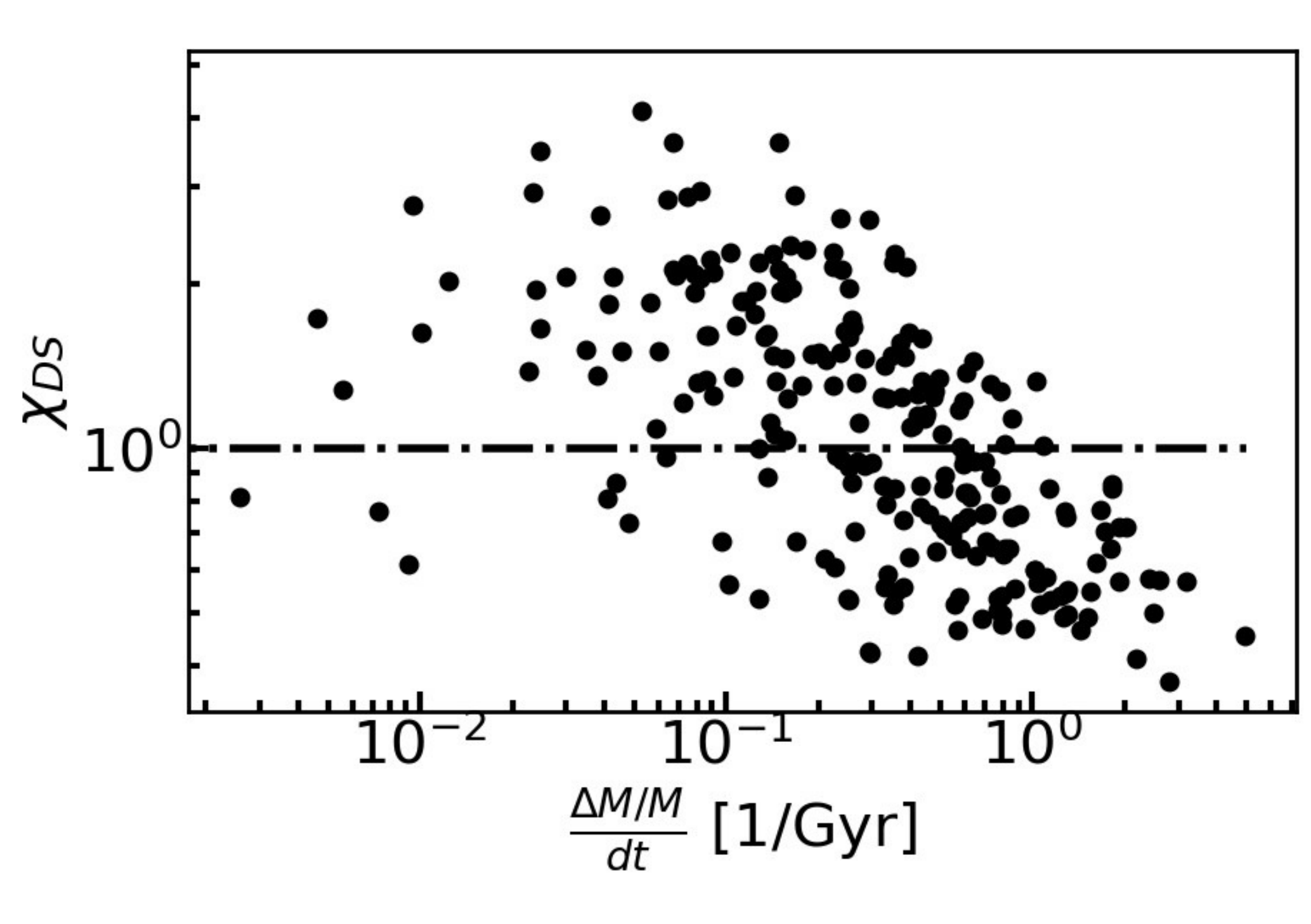}
    \caption{The relaxation parameter at $z=0.33$ is represented as a function of the mass growth from $z=0.46$ to $z=0.33$, all estimated at $\rm R_{500}$. The dot-dashed line represents the threshold under which a cluster is considered disturbed.}
 \label{fig:accretion_rate_chi}
\end{figure}

Overall Fig.\ref{fig:accretion_rate} shows that the mass growth is a parameter that is capable to distinguish between relaxed and disturbed clusters, as seen also in Fig.~\ref{fig:accretion_rate_chi}. We, indeed, find a correlation betwee the mass growth parameter and $\chi_{\rm DS}$ equal to -0.46 both at $\rm R_{500}$ (shown in the Figure) and at $\rm R_{200}$. The good correlation is actually expected since one of the indicators defining $\chi_{\rm DS}$ is the mass in substructures. That said there is a certain level of contamination with the presence of disturbed objects with reduced mass growth or viceversa relaxed objects with significant mass growth. Notice, however, that both these peculiar cases have a HE bias value which is quite consistent with the average expectation of the samples, while extreme bias values such as those with $b<-0.2$ are only present for the largest mass growth. 

\subsubsection{Relaxation parameter}
\label{subsec:relaxation_p}
\begin{figure}
    \includegraphics[width=.5\textwidth]{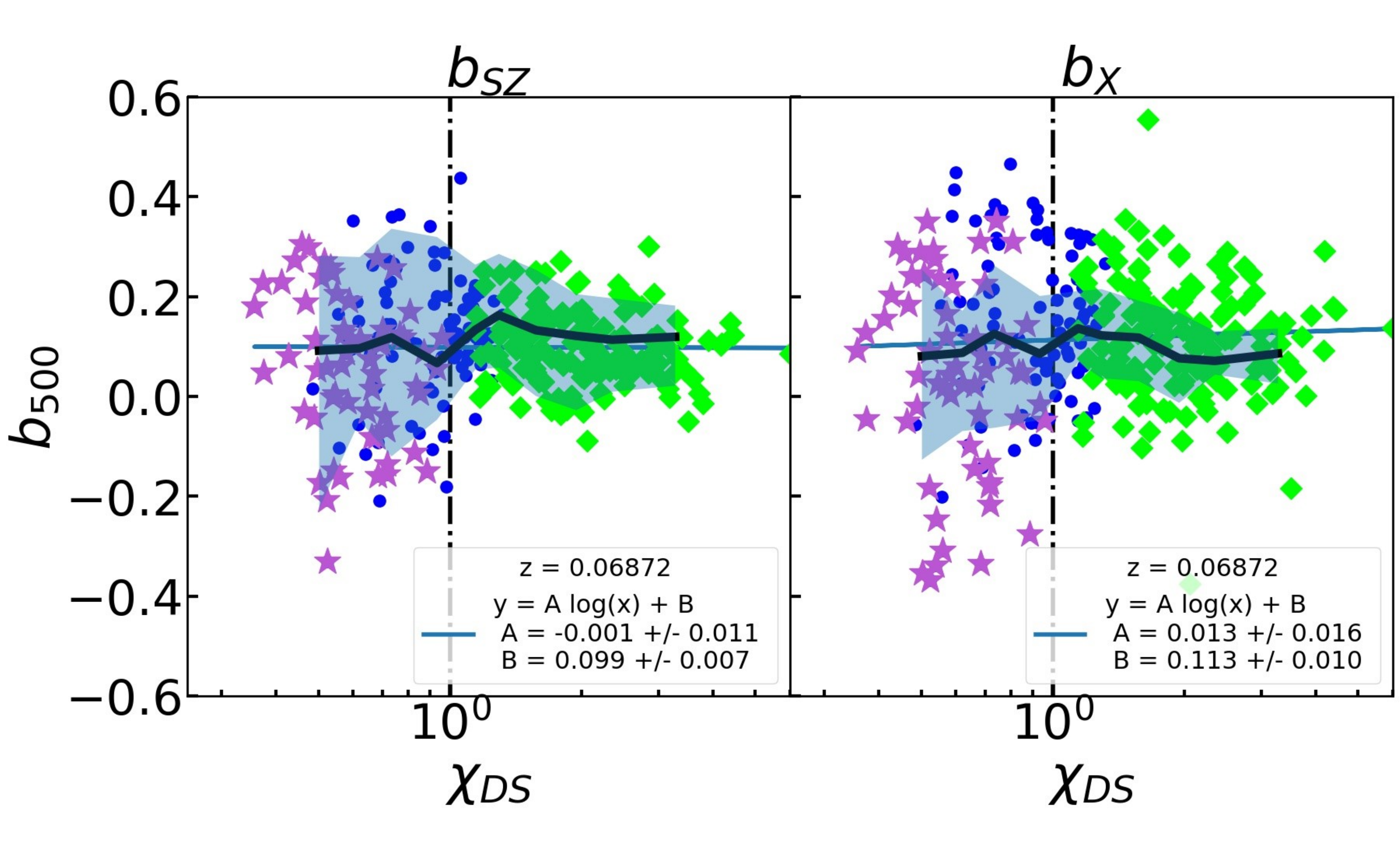}
    \includegraphics[width=.5\textwidth]{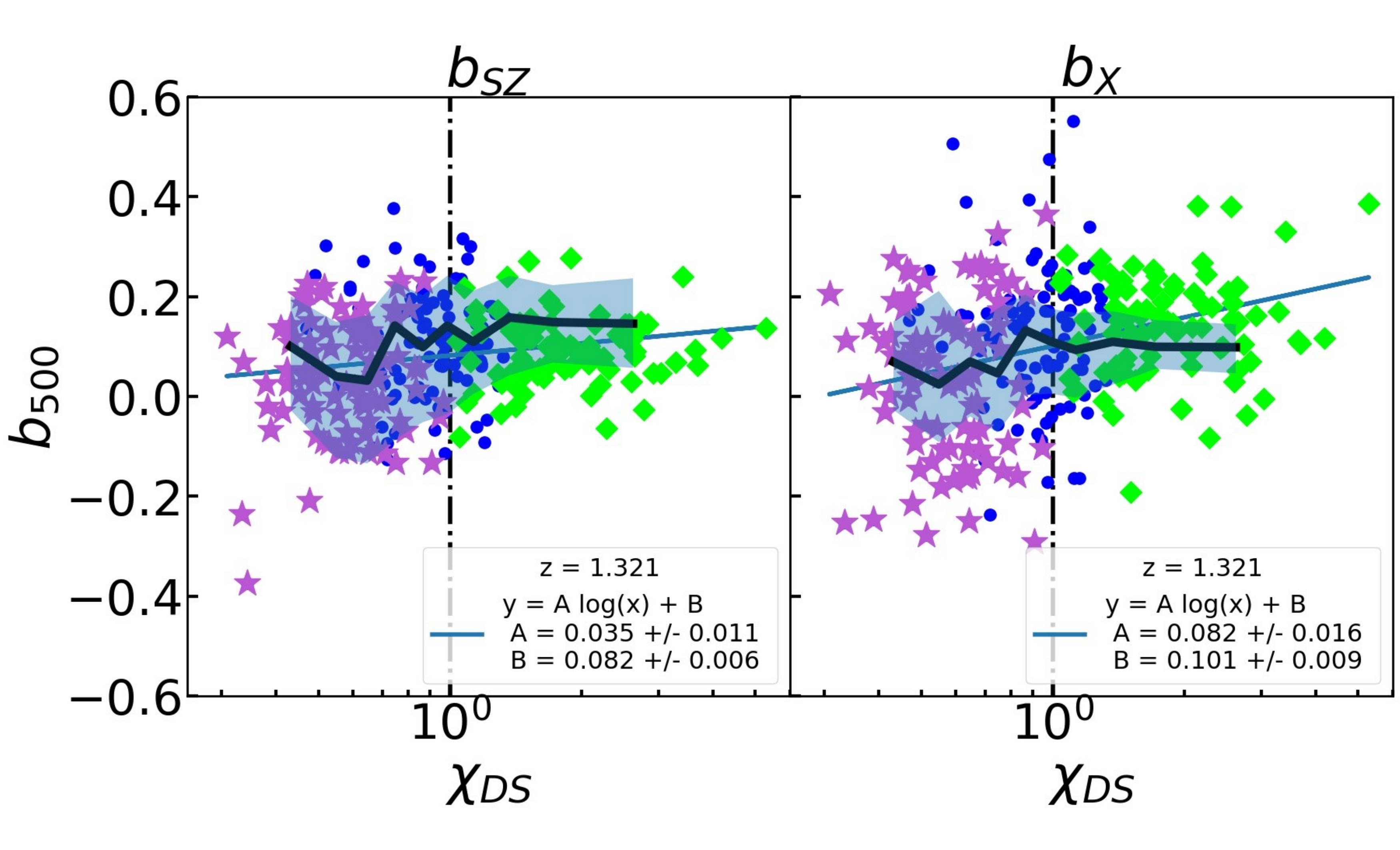}
    \caption{The biases are represented as a function of the relaxation parameter $\chi_{\rm DS}$. The hybrids, relaxed and disturbed clusters are represented with blue dots, green diamonds and purple stars respectively. In the top panel we show the quantities at $z=0.07$ and in the bottom one at $z=1.3$. The lines are the results of a simple linear fitting result. The black line and the shaded region represent the median bias and the percentiles (16th and 84th).}
 \label{fig:relaxation}
\end{figure}

The bias as a function of the relaxation parameter is represented in Fig. \ref{fig:relaxation}, where, per definition, the coloured points of relaxed and disturbed clusters are completely separated. Weak or no correlations is found, indeed the Pearson correlation coefficients with the highest values are only equal to 0.13 for $b_{\rm SZ}$ and 0.23 for $b_{\rm X}$ at $z=1.32$; and 0.01 for $b_{\rm SZ}$ and 0.01 for $b_{\rm X}$ at $z=0.07$. 
Also in this case, the disturbed clusters show a wider dispersion with respect to the relaxed ones, see Table \ref{Tab:chi}, where the difference between the bias percentiles grows by almost a factor of 2 from the lowest to the largest $\chi_{\rm DS}$ clusters. This is in agreement with \citet{piffaretti, rasia, Nelson2014, Henson2017, ansarifard}.

%

\begin{table}
\centering
\caption{Table of the bias values for the 50 highest (first row) and 50 lowest (third row) $\chi_{\rm DS}$ clusters and for the remaining clusters (second row) for the $z=0.59$ sample. We report the median value ($m$), the $16^{\rm th}$ and $84^{\rm th}$ percentiles ($p_{16}$ and $p_{84}$), and their half difference ($d_2$) at R$_{500}$.}
\begin{tabular}[t]{ccccccccc}
\hline
  $\chi_{\rm DS}$  & $b_{SZ}$ & & & & $b_X$ & \\
   & m & $p_{16}$ & $p_{84}$ & d2 & m & $p_{16}$ & $p_{84}$ & d2 \\
\hline
high & 0.10 & 0.05 & 0.16 & 0.06 & 0.12 & 0.05 & 0.23 & 0.09 \\
\hline
med & 0.11 & -0.02 & 0.19 & 0.09 & 0.14 & 0.01 & 0.25 & 0.12 \\
\hline 
low & 0.11 & -0.04 & 0.21 & 0.13 & 0.07 & -0.12 & 0.20 & 0.16 \\
\hline
\end{tabular}
\label{Tab:chi}
\end{table}

\subsubsection{Total mass}
\begin{figure}
    \includegraphics[width=.5\textwidth]{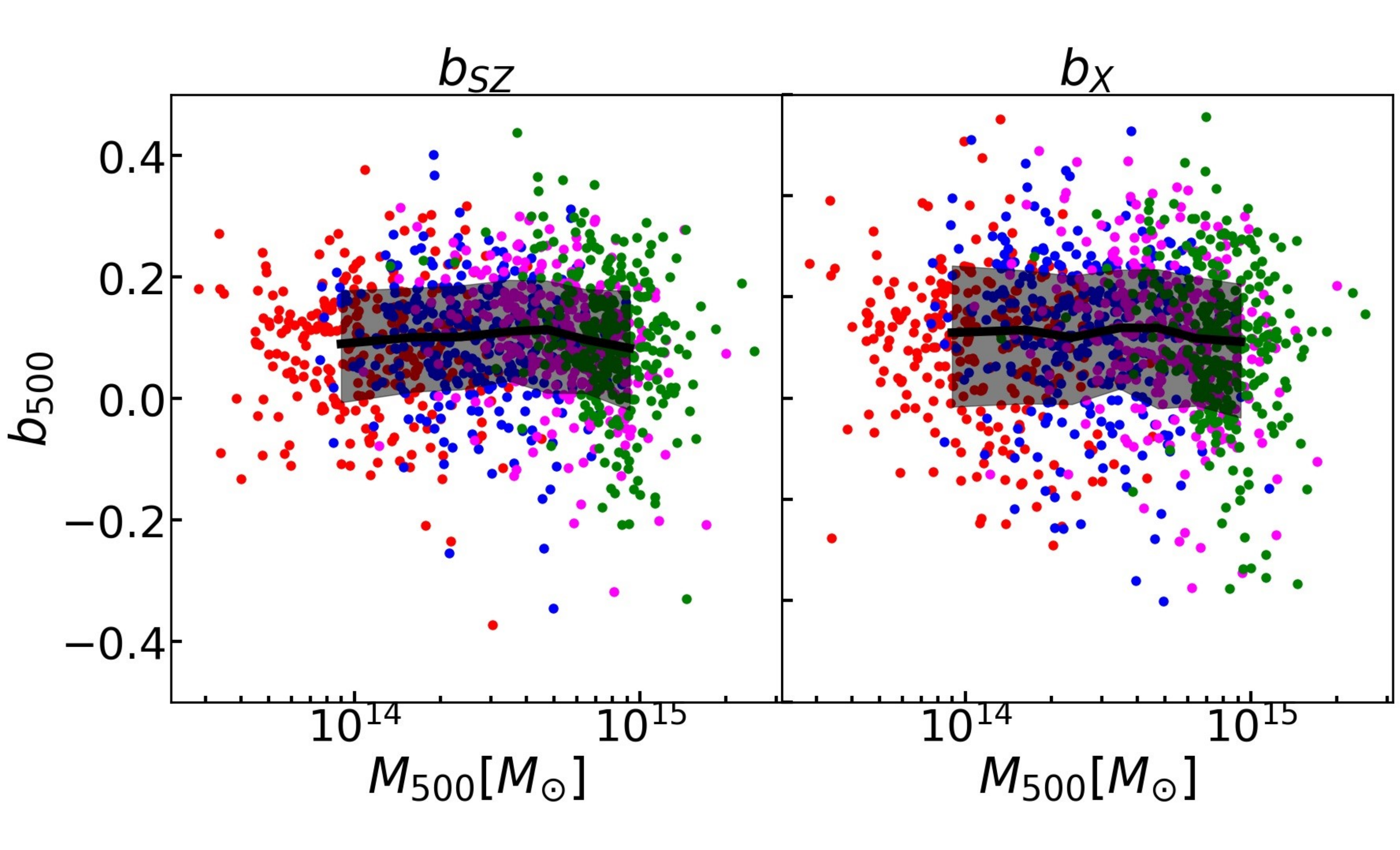}
    \caption{The biases are represented as a function of the cluster total mass, both at $\rm R_{500}$. The redshifts 1.32, 0.78, 0.33, 0.07 are represented in red, blue, magenta and green, respectively. The black line and the shaded region represent the binned median and the 16th-84th percentiles. }
 \label{fig:mass}
\end{figure}

We continue to investigate the correlation between the HE bias and the cluster mass at $\rm R_{500}$ in Fig.\ref{fig:mass}. Since we detect no variation on cosmic time (see Fig.~\ref{fig:redshift}), in this plot, we simultaneously show the biases at four redshifts (1.32, 0.78, 0.33, 0.07) to increase the halo mass range. 
We find no dependence of the bias on the total mass of the clusters in agreement with  \citet{piffaretti, lebrun, Henson2017, pearce2019, ansarifard, barnes2020}. 

\subsubsection{Triaxiality}
We study the dependence of the bias at $R_{500}$ on the sphericity of the halos, one of the HE main assumptions, quantified with the parameter $t$ and on $c/a$, the ratio of the minor and major axis. All the particles within $R_{500}$ are used to estimate the halo shape \citep{vega}. 

\begin{figure}
    \includegraphics[width=.5\textwidth]{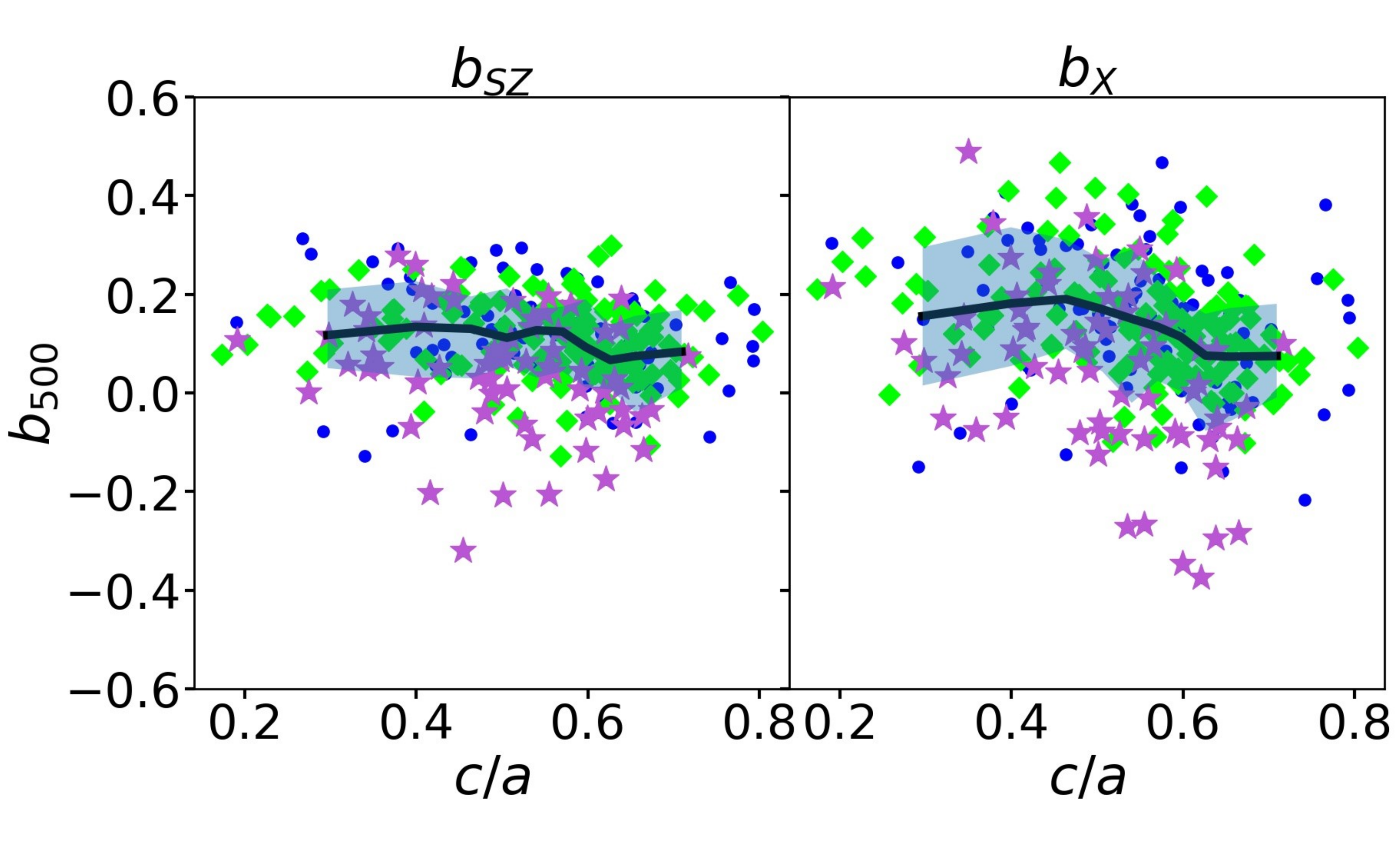}
    \caption{The biases at $\rm R_{500}$ are represented as a function of the axis ratio $c/a$. The hybrids, relaxed and disturbed are represented with blue dots, green diamonds and purple stars respectively. Both $b_{SZ}$ and $b_X$ are shown for $z=0.33$. The median value of the bias is represented with the black line, the 16th and 84th percentiles with the shaded region. }
 \label{fig:triaxiality}
\end{figure}

The relation between the biases and $c/a$ is represented in Fig. \ref{fig:triaxiality}, the results for $t$ are not shown, being very similar. From that Figure we notice that the overall correlation and the amplitude of the scatter seem to be uncorrelated with the axial ratio. Although, if we restrict to the most aspherical cases (third line in Table \ref{Tab:triaxiality}) we see that the both biases increases by almost 50\% with respect to the objects with the highest $c/a$ values.
Note that the halo shape estimated based on total mass is similar to the one based on hot gas \citep[see][for details]{Velliscig2015}. Therefore, we believe a similar result will be in place when the halo shape is estimated only from the gas particles.

\begin{table}
\centering
\caption{Table of the biases values for the 50 clusters with the highest (first row) and lowest (last row) axial ratio and for the remaining clusters (second row) for the $z=0.46$ sample.
We report the median value ($m$), the $16^{\rm th}$ and $84^{\rm th}$ percentiles ($p_{16}$ and $p_{84}$), and their half difference ($d_2$) at R$_{500}$.
}
\begin{tabular}[t]{ccccccccc}
\hline
$c/a$       & $b_{SZ}$ & & & & $b_X$ & \\
 & m & $p_{16}$ & $p_{84}$ & $d_2$ & m & $p_{16}$ & $p_{84}$ & $d_2$ \\
\hline
high & 0.09 & 0.01 & 0.18 & 0.07 & 0.11 & -0.00 & 0.18 & 0.09 \\
\hline
med & 0.09 & 0.00 & 0.17 & 0.08 & 0.14 & -0.02 & 0.23 & 0.13 \\
\hline 
low & 0.12 & 0.02 & 0.24 & 0.11 & 0.20 & 0.07 & 0.32 & 0.12 \\
\hline
\end{tabular}
\label{Tab:triaxiality}
\end{table}

%

\subsection{HE mass bias and the merger history of the clusters}
\label{sec:merger}
Unlike observations, hydrodynamical simulations allow for tracking the whole history of a cluster, {\it and} thus to follow a merging event during all its phases. Specifically in this subsection, we study major merger events that are identified when a halo experiences a very rapid mass increase, primarily caused by the accretion of a single massive object \citep{Contreras2022}. Our main goal is to understand the evolution of the bias throughout these violent merging processes.

To define and compute the relevant times to the merger event, we closely follow the definitions in \citet{Contreras2022} to which we refer for a more detailed explanations. In summary, instead of using absolute time as the variable in the merger history, which can vary with redshift, we normalise it to the halo dynamical time:
\begin{equation}
t_{dyn} = \sqrt{\frac{3}{4\pi} \frac{1}{200G\rho_{\rm crit}}} . 
\end{equation}
Since the critical density depends only on the cosmology and on the redshift, see Eq.~(\ref{eq:rho_crit}), the dynamical time does not depend on the cluster features, but evolves only with redshift. As described before, we only consider the major merger events defined as an event that produces a mass increase of 100\% within half of the dynamical time
\begin{equation}
    \frac{\Delta M}{M} = \frac{M_f - M_i}{M_i} \geq 1,
\end{equation}
where $M_i$ is the initial mass and $M_f$ is the final mass.
Note that the masses in this case refer to the overdensity of 200. We use this larger radius for better capturing the evolution process inside the virial region.
The merger event can easily be characterised by 4 particular redshifts (see Figure~1 of \citealt{Contreras2022}):
\begin{itemize}
    \item $z_{\rm before}$: the last time before the merger begins. It is characterised by a relaxation parameter with high values (typically larger than 1). Soon after the two systems start to influence each other and the relaxation parameter decreases;
    \item $z_{\rm start}$: the merging halo enters in the main cluster $R_{200}$ and as a consequence the mass of the latter starts to grow;
    \item $z_{\rm end}$: the end point of the merger identified as the moment when the mass growth stops and the $\chi_{\rm DS}$ starts to grow again implying that the relaxation process is beginning.
    \item $z_{\rm after}$: the end of the whole merger phase when the cluster approaches a relaxed state by showing $\chi_{DS}$ closer or above $1$. 
\end{itemize}

\begin{figure*}
    \includegraphics[width=0.8\textwidth]{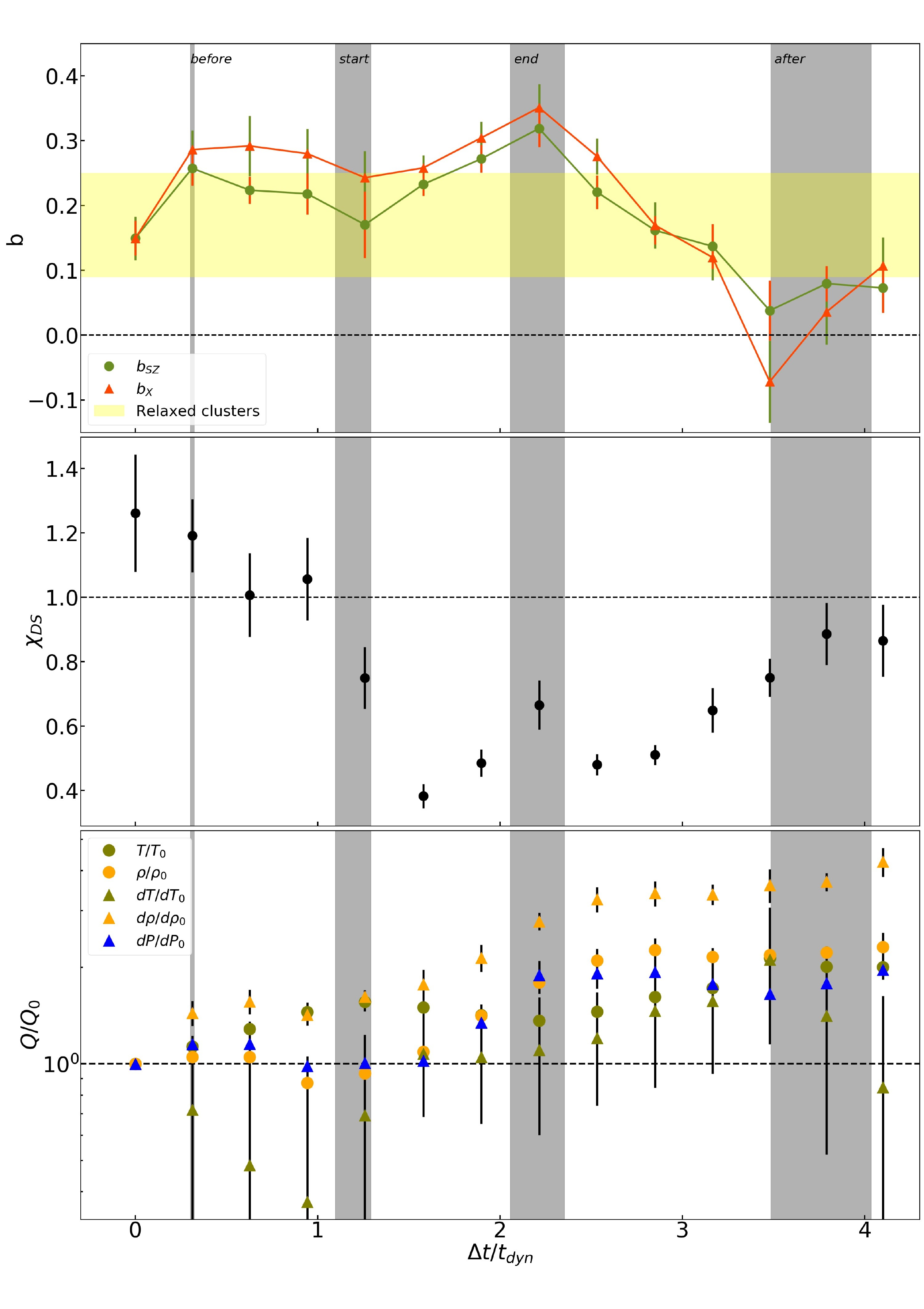}

    \caption{Top panel: the stacked biases at $\rm R_{200}$ (means and standard deviations) are represented as a function of the difference in times from $t_{0}$, divided by the dynamical time. The SZ bias is represented in green, while the X one in orange. The yellow shaded region represents the biases range of the relaxed clusters. The null bias is represented with a dashed line. Central panel: the stacked relaxation parameter estimated inside $\rm R_{200}$, the means and the standard deviations are represented as a function of $\Delta t /t_{dyn}$. The dashed line represents the threshold between relaxed and disturbed clusters. Bottom panel: each thermodynamic quantity which takes part in the estimation of the HE mass (and bias) divided by their value before the merger (indicated with a subscript 0) is shown as a function of $\Delta t/ t_{dyn}$. The dashed line shows the value of 1 (the quantity does not change with respect to the age prior to the merger). The temperature and its derivative are represented in olive green dots and triangles respectively. Instead, the density and its derivative are represented as orange dots and triangles, and the pressure derivative is represented with blue triangles. In the plot the means and the standard deviations are represented. In all the panels, the vertical grey shaded areas represent the $\pm 1\sigma$ regions of the averaged time \textit{before}, \textit{after}, \textit{start} and \textit{end} of the merger.}
 \label{fig:stacking}
\end{figure*}

In this work, we analyse how the bias evolves with the major merger event and how much time is needed for the bias {\it to return} to the average value in the relaxed population. In our sample, we select 12 clusters which experience a major merger, from $z_{\rm before}$ to $z_{\rm after}$, in the investigated redshift range [0.07,1.32].
In the top panel of Fig. \ref{fig:stacking} the two stacked biases, presented by mean bias values with $16^{th}-84^{th}$ percentiles as error bars, are shown as a function of $\Delta t / t_{\rm dyn}$, which is defined as 
\begin{equation}
    \frac{\Delta t}{t_{\rm dyn}} = \frac{t_{ 0 +\textit{i}} - t_{ 0 }}{t_{\rm dyn}} 
\label{eq:delta_t}
\end{equation}
where $t_{0}$ corresponds to the analysed redshift right before $z_{\rm before}$, 
and 
$t_{0+i}$ corresponds to the analysed redshifts following $t_0$. The vertical grey shaded areas represent the $\pm 1\sigma$ regions of the averaged time \textit{before}, \textit{after}, \textit{start} and \textit{end} of the merger. The yellow shaded region shows the $\pm1\sigma$ region of the SZ bias for the relaxed clusters at $\rm R_{200}$, averaged over all the redshifts. 

At $\Delta t / t_{\rm dyn} = 0$ both biases are in agreement with the typical relaxed clusters values. Right at the beginning of the merger phase ($z_{\rm before}$), the biases increase to $\sim 0.25-0.3$ due to the incoming substructure, that causes the true mass of the cluster to increase faster than the HE mass, leading to a slight increase of the bias. The latter stays almost constant until the end of the accretion of the secondary object. In this situation the HE mass is always underestimating the true mass of the object. A small increase of the bias is detected at around $z_{\rm end}$ followed by a quick decrease that lasts until $z_{\rm after}$, where a minimum is reached. Right at this time, around the end of the merger phase, the HE bias can even reach negative values. 
The in-fall of the substructure generates shocks propagating outward. These shocks cause a steep increase of the derivatives of the pressure, gas density and temperature profiles (see the bottom panel described later), leading to an increase of the HE mass, and a negative bias. This trend is in agreement with \citet{bennett2021} and \citet{Nelson2012}, who conducted a similar study on simulated clusters.
When the cluster approaches a relaxation phase at $\sim z_{\rm after}$, the bias returns to a value that is closer to the mean $b$ of relaxed clusters. This trend is again in agreement with what seen in the FABLE simulated clusters \citep{bennett2021}.

In the middle panel of Fig. \ref{fig:stacking} we show the evolution of the relaxation parameter during the merger process. The clusters are classified as relaxed at $t_0$ by definition, and then as the secondary cluster approaches (before $z_{\rm start}$) the $\chi_{\rm DS}$ drops. The relaxation parameter shows some fluctuation between $z_{\rm start}$ and $z_{\rm end}$ to finally grow again after the secondary objects is completely incorporated into the main cluster. At the end of the merger phase ($z_{\rm after}$), per definition $\chi_{\rm DS}$ is approaching 1. Notice that even after 4 dynamical times from the beginning of the merger the main cluster does not reach the original relaxation state, consistently with \citet{Contreras2022}. 

In the bottom panel, we shown the relative evolution of all quantities entering the HE mass equations (both thermodynamic quantities and their derivatives). To be consistent with the analysis shown in the upper panels, all the values are computed at $R_{200}$. The pressure derivative is represented with blue triangles, the temperature and its derivative are represented as olive green dots and triangles, the gas density and its derivative are represented with orange dots and triangles. 
The temperature derivative does not show any particular trend and fluctuates around the original values with large error bars. The temperature values instead grows immediately as the secondary object approaches and keeps growing until the end of the process, this is expected as it is a consequence of the shock heating produced by the merger. 

The density value does not show any particular change until the secondary structure merge into the main halo after $z_{\rm start}$. 
The slope of the gas density at $R_{200}$ also increases for the presence of the massive companion. In both cases, the variation continues up to $z_{\rm end}$ before reaching a plateau.
Finally, the pressure derivative is on the mid way between the temperature and the gas density derivatives. 
The increase in the derivative of the gas density, pressure and temperature due to the shocks generated in the merger or to the sloshing of the subclusters moving around the cluster core drive the HE masses of Eqs.~\ref{eq:Mhe_P} and \ref{eq:Mhe_T} to be closer or even above the true mass.

At $\rm R_{500}$ these considerations are still valid, the behaviour of all the quantities is the same. To conclude, the bias shows a stable evolution along the whole major merger event, albeit slightly difference between $b_{SZ}$ and $b_X$. However, the dramatic change of the bias value, especially in the period between $z_{\rm end}$ to $z_{\rm after}$ when the cluster is identified as un-relaxed, is the origin of the large scatter of the bias for the disturbed systems shown in all previous figures. For this reason, it is not suggested to estimate the HE masses for these dynamical un-relaxed clusters, unless we can correctly identify their states in the merging events which can be very difficult in observation. 

\section{Conclusions}
\label{sec:conclusions}

Galaxy cluster mass plays a major role in cluster cosmology. Several different methods are used to estimate it. In this paper we focus on the hydrostatic equilibrium approximation, which makes use of the temperature, pressure and density radial profiles of the ICM. These 3D profiles are computed from \theth\ simulations, which include a set of almost 300 simulated galaxy clusters that we studied at 10 different redshifts. From the profiles, we recover the masses under the HE approximation and we estimate the bias against the true cluster total mass. In this work the focus is specifically on the connection between the hydrostatic mass bias and different cluster properties, such as dynamical state, cluster total mass, mass-growth rate, NFW concentration and axial ratio.  
Moreover, we follow the bias during 12 major-merger events to understand the bias evolution along these events. The main findings are as follows:

\begin{itemize}
    \item We do not find correlation between the biases, estimated at different radii, and the redshift, in agreement with other simulations.
    \item The bias and its scatter are influenced by the radii within which the bias is estimated, i.e. $R_{2500}, R_{500}$ and $R_{200}$ in this work. The largest bias is at $R_{200}$ (almost 20\%), while at $R_{500}$ and $R_{2500}$, the median value is around 10\%.
    \item There is no correlation between the hydrostatic mass bias and the dynamical state of the clusters as measured with different indicators: the dynamical-state parameter $\chi_{\rm DS}$, the NFW concentration, the relative mass growth, and the triaxiality. 
    However, whenever one of these parameters is typically associated to a disturbed objects (e.g. low $\chi_{\rm DS}$, low concentration, high mass growth) we detect a bias scatter that can be almost twice as that of the relaxed systems. The scatter has, instead, little dependence on mass, redshift, and sphericity.
    \item A moderate correlation is found between the relaxation parameter and the mass growth. This is expected because the larger is the mass growth, the more likely the cluster is disturbed. 
    \item By stacking clusters while experiencing major merger events we find that the main object becomes more disturbed along the merger with a slightly higher $b$, when the secondary structure collides. After $z_{\rm end}$ -- the halo mass growth stops and $b$ drops significantly because of the shocks propagating outward generated by the in-falling substructure. These shocks cause a steep increase of the derivatives of the pressure, gas density and temperature profiles, even leading to an overestimation of the HE mass. 
    At the end of the merger phase when clusters get closer to a relaxed dynamical state with $\chi \sim 1$, the bias assumes the typical values of relaxed clusters.
\end{itemize}

Although no correlation is found between the HE bias and the dynamical state of the cluster at a given redshift, we find that selecting a sample with the typical characteristics of regular objects, such as high NFW concentration, low mass accretion rate, high $\chi_{\rm DS}$, and high $c/a$ ratio, leads to a reduced cluster-to-cluster scatter. We also found a correlation between various merger phases and the HE bias which can be close to 0 when the cluster is at the end of merger process. This, however, occurs when the cluster is not yet relaxed confirming that including yet-no relaxed clusters could largely increase the bias scatter.

Our results are in agreement with other simulations,
where a large bias is found mostly generated by  deviations from spherical symmetry (which is at the base of HE), temperature dishomogeneities, but also the presence of substructures or gas motions, generating non-thermal pressure component.
In the first part of the paper, we have proved that the HE bias is not dependent on the latter, while in the last part of the paper we explore how the bias changes during a merger. Here we show that the presence of a substructure, and consequently a merger, can indeed deeply affect the cluster dynamical state and the HE mass bias, leading even to an overestimation of the true mass. During the merger events, see Fig. 8, the bias is almost always positive, i.e. the HE mass is  underestimating the true cluster mass. However, even with the highest bias value during the merger events, we can only come close to the bias suggested by Planck. Therefore, we need to look for other reasons for explaining this bias tension. The next-generation satellites, like XRISM (X-Ray Imaging and Spectroscopy Mission), hopefully Athena and the proposed probe LEM (Line Emission Mapper), but also eROSITA-data analysis, are going to give precise indication on the gas velocities and dispersion and, consequently on cluster dynamical state.


\section*{Acknowledgements}
The authors would like to thank the Referee for the constructive and helpful comments. The simulations used in this work  have been performed
in the MareNostrum Supercomputer at the Barcelona Supercomputing Center, thanks to CPU time granted by the Red Espa\~nola  de Supercomputaci\'on. WC is supported by the STFC AGP Grant ST/V000594/1 and the Atracci\'{o}n de Talento Contract no. 2020-T1/TIC-19882 granted by the Comunidad de Madrid in Spain. He further acknowledges the science research grants from the China Manned Space Project with NO. CMS-CSST-2021-A01 and CMS-CSST-2021-B01. GY acknowledges financial support from the MICIU/FEDER (Spain) under project grant PGC2018-094975-C21. MDP acknowledges support from Sapienza Università di Roma thanks to Progetti di Ricerca Medi 2019, RM11916B7540DD8D. AK is supported by the Ministerio de Ciencia, Innovaci\'{o}n y Universidades (MICIU/FEDER) under research grant PGC2018-094975-C21 and further thanks Piero Umiliani for `svezia, inferno e paradiso'. 


\section*{Data Availability}
The data underlying this article were produced as part of \theth\ Project \citep{Cui2018}. They will be shared on request to \theth\ Collaboration, at \url{https://www.the300-project.org}.


\bibliographystyle{mnras}
\bibliography{biblio} 

\begin{thebibliography}{}
\makeatletter
\relax
\def\mn@urlcharsother{\let\do\@makeother \do\$\do\&\do\#\do\^\do\_\do\%\do\~}
\def\mn@doi{\begingroup\mn@urlcharsother \@ifnextchar [ {\mn@doi@}
  {\mn@doi@[]}}
\def\mn@doi@[#1]#2{\def\@tempa{#1}\ifx\@tempa\@empty \href
  {http://dx.doi.org/#2} {doi:#2}\else \href {http://dx.doi.org/#2} {#1}\fi
  \endgroup}
\def\mn@eprint#1#2{\mn@eprint@#1:#2::\@nil}
\def\mn@eprint@arXiv#1{\href {http://arxiv.org/abs/#1} {{\tt arXiv:#1}}}
\def\mn@eprint@dblp#1{\href {http://dblp.uni-trier.de/rec/bibtex/#1.xml}
  {dblp:#1}}
\def\mn@eprint@#1:#2:#3:#4\@nil{\def\@tempa {#1}\def\@tempb {#2}\def\@tempc
  {#3}\ifx \@tempc \@empty \let \@tempc \@tempb \let \@tempb \@tempa \fi \ifx
  \@tempb \@empty \def\@tempb {arXiv}\fi \@ifundefined
  {mn@eprint@\@tempb}{\@tempb:\@tempc}{\expandafter \expandafter \csname
  mn@eprint@\@tempb\endcsname \expandafter{\@tempc}}}

\bibitem[\protect\citeauthoryear{Angelinelli, Vazza, Giocoli, Ettori, Jones,
  Brunetti, Brüggen  \& Eckert}{Angelinelli et~al.}{2020}]{Angelinelli2020}
Angelinelli M.,  Vazza F.,  Giocoli C.,  Ettori S.,  Jones T.~W.,  Brunetti G.,
   Brüggen M.,   Eckert D.,  2020, \mn@doi [MNRAS] {10.1093/mnras/staa975},
  495, 864–885

\bibitem[\protect\citeauthoryear{Ansarifard et~al.,}{Ansarifard
  et~al.}{2020}]{ansarifard}
Ansarifard S.,  et~al., 2020, \mn@doi [\aap] {10.1051/0004-6361/201936742},
  634, A113

\bibitem[\protect\citeauthoryear{Arnaud, Pratt, Piffaretti, B{\"{o}}hringer,
  Croston  \& Pointecouteau}{Arnaud et~al.}{2010}]{arnaud10}
Arnaud M.,  Pratt G.~W.,  Piffaretti R.,  B{\"{o}}hringer H.,  Croston J.~H.,
  Pointecouteau E.,  2010, \mn@doi [\aap] {10.1051/0004-6361/200913416}, 517, 1

\bibitem[\protect\citeauthoryear{Artis, Melin, Bartlett  \& Murray}{Artis
  et~al.}{2022}]{Artis2021}
Artis E.,  Melin J.-B.,  Bartlett J.,   Murray C.,  2022, \mn@doi [EPJ Web of
  Conferences] {10.1051/epjconf/202225700004}, 257, 00004

\bibitem[\protect\citeauthoryear{Barnes, Vogelsberger, Pearce, Pop, Kannan,
  Cao, Kay  \& Hernquist}{Barnes et~al.}{2021}]{barnes2020}
Barnes D.~J.,  Vogelsberger M.,  Pearce F.~A.,  Pop A.-R.,  Kannan R.,  Cao K.,
   Kay S.~T.,   Hernquist L.,  2021, \mn@doi [\mnras] {10.1093/mnras/stab1276},
  506

\bibitem[\protect\citeauthoryear{Bennett \& Sijacki}{Bennett \&
  Sijacki}{2022}]{bennett2021}
Bennett J.~S.,  Sijacki D.,  2022, \mn@doi [MNRAS] {10.1093/mnras/stac1216},
  514, 313

\bibitem[\protect\citeauthoryear{Biffi, Sembolini, {De Petris}, Valdarnini,
  Yepes  \& Gottl{\"{o}}ber}{Biffi et~al.}{2014}]{Biffi2014}
Biffi V.,  Sembolini F.,  {De Petris} M.,  Valdarnini R.,  Yepes G.,
  Gottl{\"{o}}ber S.,  2014, \mn@doi [\mnras] {10.1093/mnras/stu018}, 439, 588

\bibitem[\protect\citeauthoryear{Bulbul et~al.,}{Bulbul
  et~al.}{2019}]{Bulbul2019}
Bulbul E.,  et~al., 2019, \mn@doi [ApJ] {10.3847/1538-4357/aaf230}, 871, 50

\bibitem[\protect\citeauthoryear{Campitiello et~al.,}{Campitiello
  et~al.}{2022}]{campitiello2022}
Campitiello M.~G.,  et~al., 2022, \mn@doi [A\&A] {10.1051/0004-6361/202243470},
  665, A117

\bibitem[\protect\citeauthoryear{Cialone, {De Petris}, Sembolini, Yepes, Baldi
  \& Rasia}{Cialone et~al.}{2018}]{Cialone}
Cialone G.,  {De Petris} M.,  Sembolini F.,  Yepes G.,  Baldi A.~S.,   Rasia
  E.,  2018, \mn@doi [\mnras] {10.1093/mnras/sty621}, 477, 139

\bibitem[\protect\citeauthoryear{Contreras-Santos et~al.,}{Contreras-Santos
  et~al.}{2022}]{Contreras2022}
Contreras-Santos A.,  et~al., 2022, \mn@doi [MNRAS] {10.1093/mnras/stac275},
  511, 2897

\bibitem[\protect\citeauthoryear{{Cui} et~al.,}{{Cui} et~al.}{2016}]{Cui2016}
{Cui} W.,  et~al., 2016, \mn@doi [\mnras] {10.1093/mnras/stv2839}, \href
  {https://ui.adsabs.harvard.edu/abs/2016MNRAS.456.2566C} {456, 2566}

\bibitem[\protect\citeauthoryear{{Cui}, {Power}, {Borgani}, {Knebe}, {Lewis},
  {Murante}  \& {Poole}}{{Cui} et~al.}{2017}]{cui2017}
{Cui} W.,  {Power} C.,  {Borgani} S.,  {Knebe} A.,  {Lewis} G.~F.,  {Murante}
  G.,   {Poole} G.~B.,  2017, \mn@doi [\mnras] {10.1093/mnras/stw2567}, \href
  {https://ui.adsabs.harvard.edu/abs/2017MNRAS.464.2502C} {464, 2502}

\bibitem[\protect\citeauthoryear{Cui et~al.,}{Cui et~al.}{2018}]{Cui2018}
Cui W.,  et~al., 2018, \mn@doi [MNRAS] {10.1093/mnras/sty2111}, 480, 2898

\bibitem[\protect\citeauthoryear{Cui et~al.,}{Cui et~al.}{2022}]{Cui2022}
Cui W.,  et~al., 2022, \mn@doi [MNRAS] {10.1093/mnras/stac1402}, 514, 977

\bibitem[\protect\citeauthoryear{Davé, Anglés-Alcázar, Narayanan, Li,
  Rafieferantsoa  \& Appleby}{Davé et~al.}{2019}]{Dave2019}
Davé R.,  Anglés-Alcázar D.,  Narayanan D.,  Li Q.,  Rafieferantsoa M.~H.,
  Appleby S.,  2019, \mn@doi [MNRAS] {10.1093/mnras/stz937}, 486, 2827

\bibitem[\protect\citeauthoryear{De~Andres et~al.,}{De~Andres
  et~al.}{2022}]{deandres2021}
De~Andres D.,  et~al., 2022, \mn@doi [EPJ Web of Conferences]
  {10.1051/epjconf/202225700013}, 257, 00013

\bibitem[\protect\citeauthoryear{De Luca, De Petris, Yepes, Cui, Knebe  \&
  Rasia}{De Luca et~al.}{2021}]{DeLuca}
De Luca F.,  De Petris M.,  Yepes G.,  Cui W.,  Knebe A.,   Rasia E.,  2021,
  \mn@doi [MNRAS] {10.1093/mnras/stab1073}, 504, 5383

\bibitem[\protect\citeauthoryear{Ettori, Donnarumma, Pointecouteau, Reiprich,
  Giodini, Lovisari  \& Schmidt}{Ettori et~al.}{2013}]{bib:ettori_rev}
Ettori S.,  Donnarumma A.,  Pointecouteau E.,  Reiprich H.~T.,  Giodini S.,
  Lovisari L.,   Schmidt R.~W.,  2013, \mn@doi [Space Sci Rev]
  {10.1007/s11214-013-9976-7}, 177, 119–154

\bibitem[\protect\citeauthoryear{Gianfagna et~al.,}{Gianfagna
  et~al.}{2021}]{gianfagna2021}
Gianfagna G.,  et~al., 2021, \mn@doi [MNRAS] {10.1093/mnras/stab308}, 502, 5115

\bibitem[\protect\citeauthoryear{Haggar, Gray, Pearce, Knebe, Cui, Mostoghiu
  \& Yepes}{Haggar et~al.}{2020}]{Haggar2020}
Haggar R.,  Gray M.~E.,  Pearce F.~R.,  Knebe A.,  Cui W.,  Mostoghiu R.,
  Yepes G.,  2020, \mn@doi [\mnras] {10.1093/mnras/staa273}, 492, 6074

\bibitem[\protect\citeauthoryear{Henson, Barnes, Kay, McCarthy  \&
  Schaye}{Henson et~al.}{2017}]{Henson2017}
Henson M.~A.,  Barnes D.~J.,  Kay S.~T.,  McCarthy I.~G.,   Schaye J.,  2017,
  \mn@doi [\mnras] {10.1093/mnras/stw2899}, 465, 3361

\bibitem[\protect\citeauthoryear{Hern{\'{a} }ndez-Lang et~al.,}{Hern{\'{a}
  }ndez-Lang et~al.}{2022}]{hernandezlang2021}
Hern{\'{a} }ndez-Lang D.,  et~al., 2022, \mn@doi [MNRAS]
  {10.1093/mnras/stac2480}, 517, 4355

\bibitem[\protect\citeauthoryear{{Hoekstra}, {Herbonnet}, {Muzzin}, {Babul},
  {Mahdavi}, {Viola}  \& {Cacciato}}{{Hoekstra} et~al.}{2015}]{Hoekstra2015}
{Hoekstra} H.,  {Herbonnet} R.,  {Muzzin} A.,  {Babul} A.,  {Mahdavi} A.,
  {Viola} M.,   {Cacciato} M.,  2015, \mn@doi [\mnras] {10.1093/mnras/stv275},
  \href {https://ui.adsabs.harvard.edu/abs/2015MNRAS.449..685H} {449, 685}

\bibitem[\protect\citeauthoryear{Klypin, Yepes, Gottlöber, Prada  \&
  Heß}{Klypin et~al.}{2016}]{Klypin2016}
Klypin A.,  Yepes G.,  Gottlöber S.,  Prada F.,   Heß S.,  2016, \mn@doi
  [MNRAS] {10.1093/mnras/stw248}, 457, 4340

\bibitem[\protect\citeauthoryear{Koukoufilippas, Alonso, Bilicki  \&
  Peacock}{Koukoufilippas et~al.}{2020}]{Koukoufilippas2020}
Koukoufilippas N.,  Alonso D.,  Bilicki M.,   Peacock J.~A.,  2020, \mn@doi
  [\mnras] {10.1093/mnras/stz3351}, 491, 5464

\bibitem[\protect\citeauthoryear{Kravtsov \& Borgani}{Kravtsov \&
  Borgani}{2012}]{bib:kravtsov_rev}
Kravtsov A.~V.,  Borgani S.,  2012, \mn@doi [\araa]
  {10.1146/annurev-astro-081811-125502}, 50, 353

\bibitem[\protect\citeauthoryear{{Le Brun}, McCarthy, Schaye  \& Ponman}{{Le
  Brun} et~al.}{2017}]{lebrun}
{Le Brun} A. M.~C.,  McCarthy I.~G.,  Schaye J.,   Ponman T.~J.,  2017, \mn@doi
  [\mnras] {10.1093/mnras/stw3361}, 466, 4442

\bibitem[\protect\citeauthoryear{{Li}, {Han}, {Wang}, {Cui}, {Li}  \&
  {Yang}}{{Li} et~al.}{2021}]{Li2021}
{Li} Q.,  {Han} J.,  {Wang} W.,  {Cui} W.,  {Li} Z.,   {Yang} X.,  2021,
  \mn@doi [\mnras] {10.1093/mnras/stab1633}, \href
  {https://ui.adsabs.harvard.edu/abs/2021MNRAS.505.3907L} {505, 3907}

\bibitem[\protect\citeauthoryear{Mazzotta, Rasia, Moscardini  \&
  Tormen}{Mazzotta et~al.}{2004}]{mazzotta2004}
Mazzotta P.,  Rasia E.,  Moscardini L.,   Tormen G.,  2004, \mn@doi [\mnras]
  {10.1111/j.1365-2966.2004.08167.x}, 354, 10

\bibitem[\protect\citeauthoryear{Nagai, Kravtsov  \& Vikhlinin}{Nagai
  et~al.}{2007}]{nagai}
Nagai D.,  Kravtsov A.~V.,   Vikhlinin A.,  2007, \mn@doi [\apj]
  {10.1086/521328}, 668, 1

\bibitem[\protect\citeauthoryear{Nagarajan et~al.,}{Nagarajan
  et~al.}{2018}]{Nagarajan2018}
Nagarajan A.,  et~al., 2018, \mn@doi [MNRAS] {10.1093/mnras/sty1904}, 488, 1728

\bibitem[\protect\citeauthoryear{Navarro, Frenk  \& White}{Navarro
  et~al.}{1997}]{nfw}
Navarro J.~F.,  Frenk C.~S.,   White S. D.~M.,  1997, \mn@doi [ApJ]
  {10.1086/304888}, 490, 493

\bibitem[\protect\citeauthoryear{Nelson, Rudd, Shaw  \& Nagai}{Nelson
  et~al.}{2012}]{Nelson2012}
Nelson K.,  Rudd D.~H.,  Shaw L.,   Nagai D.,  2012, \mn@doi [ApJ]
  {10.1088/0004-637x/751/2/121}, 751, 121

\bibitem[\protect\citeauthoryear{{Nelson}, {Lau}, {Nagai}, {Rudd}  \&
  {Yu}}{{Nelson} et~al.}{2014}]{Nelson2014}
{Nelson} K.,  {Lau} E.~T.,  {Nagai} D.,  {Rudd} D.~H.,   {Yu} L.,  2014,
  \mn@doi [\apj] {10.1088/0004-637X/782/2/107}, \href
  {https://ui.adsabs.harvard.edu/abs/2014ApJ...782..107N} {782, 107}

\bibitem[\protect\citeauthoryear{Neto et~al.,}{Neto et~al.}{2007}]{neto}
Neto A.~F.,  et~al., 2007, \mn@doi [MNRAS] {10.1111/j.1365-2966.2007.12381.x},
  381, 1450

\bibitem[\protect\citeauthoryear{{Ntampaka}, {Trac}, {Sutherland}, {Battaglia},
  {P{\'o}czos}  \& {Schneider}}{{Ntampaka} et~al.}{2015}]{Ntampaka2015}
{Ntampaka} M.,  {Trac} H.,  {Sutherland} D.~J.,  {Battaglia} N.,  {P{\'o}czos}
  B.,   {Schneider} J.,  2015, \mn@doi [\apj] {10.1088/0004-637X/803/2/50},
  \href {https://ui.adsabs.harvard.edu/abs/2015ApJ...803...50N} {803, 50}

\bibitem[\protect\citeauthoryear{{Okabe} \& {Smith}}{{Okabe} \&
  {Smith}}{2016}]{Okabe2016}
{Okabe} N.,  {Smith} G.~P.,  2016, \mn@doi [\mnras] {10.1093/mnras/stw1539},
  \href {https://ui.adsabs.harvard.edu/abs/2016MNRAS.461.3794O} {461, 3794}

\bibitem[\protect\citeauthoryear{Pearce, Kay, Barnes, Bower  \&
  Schaller}{Pearce et~al.}{2019}]{pearce2019}
Pearce F.~A.,  Kay S.~T.,  Barnes D.~J.,  Bower R.~G.,   Schaller M.,  2019,
  \mn@doi [MNRAS] {10.1093/mnras/stz3003}, 491, 1622

\bibitem[\protect\citeauthoryear{Piffaretti \& Valdarnini}{Piffaretti \&
  Valdarnini}{2008}]{piffaretti}
Piffaretti R.,  Valdarnini R.,  2008, \mn@doi [\aap]
  {10.1051/0004-6361:200809739}, 491, 71

\bibitem[\protect\citeauthoryear{{Planck Collaboration} et~al.,}{{Planck
  Collaboration} et~al.}{2016}]{Planck2016}
{Planck Collaboration} et~al., 2016, \mn@doi [A\&A]
  {10.1051/0004-6361/201525830}, 594, A13

\bibitem[\protect\citeauthoryear{Pratt, Arnaud, Biviano, Eckert, Ettori, Nagai,
  Okabe  \& Reiprich}{Pratt et~al.}{2019}]{Pratt2019}
Pratt G.~W.,  Arnaud M.,  Biviano A.,  Eckert D.,  Ettori S.,  Nagai D.,  Okabe
  N.,   Reiprich T.~H.,  2019, \mn@doi [Space Sci Rev]
  {10.1007/s11214-019-0591-0}, 215

\bibitem[\protect\citeauthoryear{Rasia et~al.,}{Rasia et~al.}{2012}]{rasia}
Rasia E.,  et~al., 2012, \mn@doi [New Journal of Physics]
  {10.1088/1367-2630/14/5/055018}, 14, 055018

\bibitem[\protect\citeauthoryear{Rasia, Borgani, Ettori, Mazzotta  \&
  Meneghetti}{Rasia et~al.}{2013}]{Rasia2013}
Rasia E.,  Borgani S.,  Ettori S.,  Mazzotta P.,   Meneghetti M.,  2013,
  \mn@doi [ApJ] {10.1088/0004-637x/776/1/39}, 776, 39

\bibitem[\protect\citeauthoryear{Rasia et~al.,}{Rasia et~al.}{2015}]{Rasia2015}
Rasia E.,  et~al., 2015, \mn@doi [ApJ] {10.1088/2041-8205/813/1/l17}, 813, L17

\bibitem[\protect\citeauthoryear{Salvati, Douspis  \& Aghanim}{Salvati
  et~al.}{2018}]{salvati}
Salvati L.,  Douspis M.,   Aghanim N.,  2018, \mn@doi [\aap]
  {10.1051/0004-6361/201731990}, 614, A13

\bibitem[\protect\citeauthoryear{Salvati, Douspis, Ritz, Aghanim  \&
  Babul}{Salvati et~al.}{2019}]{Salvati2019}
Salvati L.,  Douspis M.,  Ritz A.,  Aghanim N.,   Babul A.,  2019, \mn@doi
  [A\&A] {10.1051/0004-6361/201935041}, 626, A27

\bibitem[\protect\citeauthoryear{Sembolini, Yepes, De~Petris, Gottlöber,
  Lamagna  \& Comis}{Sembolini et~al.}{2013}]{Sembolini2013}
Sembolini F.,  Yepes G.,  De~Petris M.,  Gottlöber S.,  Lamagna L.,   Comis
  B.,  2013, \mn@doi [MNRAS] {10.1093/mnras/sts339}, 429, 323

\bibitem[\protect\citeauthoryear{Sereno, Lovisari, Cui  \&
  Schellenberger}{Sereno et~al.}{2021}]{Sereno2021}
Sereno M.,  Lovisari L.,  Cui W.,   Schellenberger G.,  2021, \mn@doi [MNRAS]
  {10.1093/mnras/stab2435}, 507, 5214–5223

\bibitem[\protect\citeauthoryear{Tian, Cheng, McGaugh, Ko  \& Hsu}{Tian
  et~al.}{2021}]{Tian2021}
Tian Y.,  Cheng H.,  McGaugh S.~S.,  Ko C.-M.,   Hsu Y.-H.,  2021, \mn@doi [ApJ
  Letters] {10.3847/2041-8213/ac1a18}, 917, L24

\bibitem[\protect\citeauthoryear{Vega-Ferrero, Yepes  \&
  Gottlöber}{Vega-Ferrero et~al.}{2017}]{vega}
Vega-Ferrero J.,  Yepes G.,   Gottlöber S.,  2017, \mn@doi [MNRAS]
  {10.1093/mnras/stx282}, 467, 3226

\bibitem[\protect\citeauthoryear{{Velliscig} et~al.,}{{Velliscig}
  et~al.}{2015}]{Velliscig2015}
{Velliscig} M.,  et~al., 2015, \mn@doi [\mnras] {10.1093/mnras/stv1690}, \href
  {https://ui.adsabs.harvard.edu/abs/2015MNRAS.453..721V} {453, 721}

\bibitem[\protect\citeauthoryear{Vikhlinin, Kravtsov, Forman, Jones,
  Markevitch, Murray  \& Speybroeck}{Vikhlinin et~al.}{2006}]{vikh}
Vikhlinin A.,  Kravtsov A.,  Forman W.,  Jones C.,  Markevitch M.,  Murray
  S.~S.,   Speybroeck L.~V.,  2006, \mn@doi [\apj] {10.1086/500288}, 640, 691

\bibitem[\protect\citeauthoryear{Von~der Linden et~al.,}{Von~der Linden
  et~al.}{2014}]{vonderLinden2014}
Von~der Linden A.,  et~al., 2014, \mn@doi [MNRAS] {10.1093/mnras/stu1423}, 443,
  1973–1978

\bibitem[\protect\citeauthoryear{Wicker, Douspis, Salvati  \& Aghanim}{Wicker
  et~al.}{2022}]{wicker2021}
Wicker R.,  Douspis M.,  Salvati L.,   Aghanim N.,  2022, \mn@doi [EPJ Web of
  Conferences] {10.1051/epjconf/202225700046}, 257, 00046

\bibitem[\protect\citeauthoryear{{Zwicky}}{{Zwicky}}{1937}]{Zwicky1937}
{Zwicky} F.,  1937, \mn@doi [\apj] {10.1086/143864}, \href
  {https://ui.adsabs.harvard.edu/abs/1937ApJ....86..217Z} {86, 217}

\makeatother
\end{thebibliography}





\bsp	
\label{lastpage}
\end{document}